\documentclass{article}

 \PassOptionsToPackage{numbers, compress}{natbib}

\usepackage[preprint]{neurips_2023}
\usepackage[utf8]{inputenc} % allow utf-8 input
\usepackage[T1]{fontenc}    % use 8-bit T1 fonts
\usepackage{hyperref}       % hyperlinks
\usepackage{url}            % simple URL typesetting
\usepackage{booktabs}       % professional-quality tables
\usepackage{amsfonts}       % blackboard math symbols
\usepackage{nicefrac}       % compact symbols for 1/2, etc.
\usepackage{microtype}      % microtypography
\usepackage{xcolor}         % colors

\usepackage{wrapfig}
\usepackage{bm}
\usepackage{amsmath}
\usepackage{amssymb}
\usepackage{amsthm}
\usepackage{amsmath}
\usepackage{graphicx}
\usepackage{subfigure}
\usepackage{enumitem}
\usepackage{diagbox}
\usepackage{multirow}
\usepackage{apptools}
\usepackage{diagbox}
\usepackage{makecell}

\usepackage{graphicx}
\usepackage{algorithm}
\usepackage[noend]{algpseudocode}

\newtheorem{theorem}{Theorem}

\newtheorem{corollary}{Corollary}[theorem]
\newtheorem{proposition}{Proposition}
\newtheorem{conjecture}{Conjecture}
\newtheorem{definition}{Definition}

\newtheorem{remark}{Remark}

\definecolor{azure}{rgb}{0.0, 0.5, 1.0}
\definecolor{brandeisblue}{rgb}{0.0, 0.44, 1.0}
\definecolor{darkpastelgreen}{rgb}{0.01, 0.75, 0.24}
\definecolor{darkpastelpurple}{rgb}{0.59, 0.44, 0.84}
\definecolor{darktangerine}{rgb}{1.0, 0.66, 0.07}
\definecolor{debianred}{rgb}{0.84, 0.04, 0.33}
\definecolor{hanpurple}{rgb}{0.32, 0.09, 0.98}
\definecolor{deepcerise}{rgb}{0.85, 0.2, 0.53}
\definecolor{emerald}{rgb}{0.31, 0.78, 0.47}
\definecolor{fuchsia}{rgb}{1.0, 0.0, 1.0}
\definecolor{flamingopink}{rgb}{0.99, 0.56, 0.67}
\definecolor{lightseagreen}{rgb}{0.13, 0.7, 0.67}
\definecolor{mayablue}{rgb}{0.45, 0.76, 0.98}

\newcommand{\EE}{\mathbb{E}}

\newcommand{\M}{\mathcal{M}}
\newcommand{\C}{\mathcal{C}}
\newcommand{\LL}{\mathcal{L}}

\newcommand{\R}{\mathbb{R}}
\newcommand{\DD}{\mathcal{D}}

\newcommand{\Ss}{\mathcal{S}}
\newcommand{\T}{\mathcal{T}}
\newcommand{\OO}{\mathcal{O}}

\DeclareMathOperator*{\arctanh}{arctanh}
\DeclareMathOperator*{\sign}{sign}
\DeclareMathOperator*{\argmax}{arg\,max}
\DeclareMathOperator*{\argmin}{arg\,min}

\DeclareMathOperator*{\bce}{bce}

\DeclareMathOperator*{\logit}{logit}
\DeclareMathOperator*{\id}{Id}
\DeclareMathOperator*{\elo}{Elo}
\DeclareMathOperator*{\rank}{rank}

\usepackage{tcolorbox}

\usepackage{todonotes}

\title{Ordinal Potential-based Player Rating}

\author{
  Nelson Vadori \\
  J.P. Morgan AI Research\\
  \And
  Rahul Savani \\
  University of Liverpool\\
  and\\
  The Alan Turing Institute
  \normalsize
}

\begin{document}

\maketitle

\begin{abstract}
It was recently observed that Elo ratings fail at preserving transitive relations among strategies and therefore cannot correctly extract the transitive component of a game. We provide a characterization of transitive games as a weak variant of ordinal potential games and show that Elo ratings actually do preserve transitivity when computed in the right space, using suitable invertible mappings. Leveraging this insight, we introduce a new game decomposition of an arbitrary game into transitive and cyclic components that is learnt using a neural network-based architecture and that prioritises capturing the sign pattern of the game, namely transitive and cyclic relations among strategies. We link our approach to the known concept of sign-rank, and evaluate our methodology using both toy examples and empirical data from real-world games.
\end{abstract}

\section{INTRODUCTION}

The Elo rating system, proposed in 1961~\cite{elo1}, assigns
ratings to players in competitive games.
Originally developed for chess, it is also widely used
across other sports (Basketball, Pool), board games, (Go,
Backgammon), and e-sports (League of Legends, StartCraft II).
Within a given pool of players, a player rating serves as a measure
of the player's relative skill within the pool,
with the probability estimate of one player beating another given as 
the sigmoid function applied to the difference in their Elo ratings.

%\noindent
%\textbf{Transitivity of games.}
%
As is common in the literature~\cite{2pszs,BalduzziTPG18,BWG23},
we formalize this problem as that of assigning ratings to the pure strategies of
a two-player symmetric zero-sum \emph{meta game}, where each pure strategy of 
the meta game corresponds to one of the players we would like to rank~\cite{2pszs}.
Such a game is called \emph{transitive} if for any pure strategies $x$, $y$,
$z$, if $x$ beats $y$, and $y$ beats $z$, then $x$ beats $z$. 
By contrast, rock-paper-scissors,
where paper beats rock, scissors beats paper, but scissors loses to rock, is
\emph{cyclic}.

Games can be transitive, cyclic, or \emph{hybrid}. 
Real-world games tend to be hybrid, with both transitive and cyclic components.
For example, \cite{CzarneckiGTTOBJ20} show that a wide range of real-world games 
are well represented by a ``spinning
top'': the upright axis represents transitive strength (i.e., the skill level of players), 
and the radial axis represents the number of cycles that exist at a particular skill level;
there are many cycles at medium
skill levels, few cycles for low skill levels, and fewer still for high skill levels.
Elo ratings are based on the assumption that the game has a significant
transitive component\footnote{In a cyclic game, no meaningful distinct skill levels can be
assigned to the pure strategies.}.
The level of transitivity of a game has been found to significantly impact which
methods are effective for training agents in these games.
For example, it has been observed that self-play struggles if the game does not 
have a suitably strong transitive component~\cite{2pszs,CzarneckiGTTOBJ20}. 
Consequently, research has focused on understanding the transitive and
cyclic components of hybrid games, e.g., through game decompositions, and the
related problem of rating players in such games~\cite{2pszs,BalduzziTPG18,BWG23,CzarneckiGTTOBJ20}.

%\noindent
%\textbf{Decompositions of games into transitive and cyclic parts.}
%%%%%%%%%%%%%%%%%%%%%%%%%%%%%%%%%%%%%%%%%%%%%%%%%%%%%%%%%%%%%%%%%%%%%%%%%%%%%%%
% Mutldimensional Elo - BalduzziTPG18
%%%%%%%%%%%%%%%%%%%%%%%%%%%%%%%%%%%%%%%%%%%%%%%%%%%%%%%%%%%%%%%%%%%%%%%%%%%%%%%
\cite{BalduzziTPG18} proposed $m$-Elo (for multidimensional Elo), which extends
the Elo score and can express cyclic components;
the same approach was independently taken by~\cite{StrangAT22}.
Using the idea of Hodge decomposition from~\cite{JiangLYY11,dgame}),
this approach first imposes a transitive component corresponding to Elo scores 
and then applies the normal (Schur) decomposition to the residual antisymmetric 
matrix after subtracting the transitive component.
%that results from subtracting the transitive component the original game.
%%%%%%%%%%%%%%%%%%%%%%%%%%%%%%%%%%%%%%%%%%%%%%%%%%%%%%%%%%%%%%%%%%%%%%%%%%%%%%%
% Disc rating - BWG23
%%%%%%%%%%%%%%%%%%%%%%%%%%%%%%%%%%%%%%%%%%%%%%%%%%%%%%%%%%%%%%%%%%%%%%%%%%%%%%%
In a more recent paper, \cite{BWG23} also use normal decomposition, but do not
impose a transitive component.
They show that their decomposition has an intuitive interpretation: each
component is a transitive or cyclic disc game.
Moreover, they show that their decomposition will contain at most one transitive component
(but possibly many cyclic components).
They use the decomposition to create a ``disc rating'' system, where each player
gets not one but two scores: skill and consistency.
%
%%%%%%%%%%%%%%%%%%%%%%%%%%%%%%%%%%%%%%%%%%%%%%%%%%%%%%%%%%%%%%%%%%%%%%%%%%%%%%%
% Balduzzi2019 -- \cite{2pszs}
%%%%%%%%%%%%%%%%%%%%%%%%%%%%%%%%%%%%%%%%%%%%%%%%%%%%%%%%%%%%%%%%%%%%%%%%%%%%%%%
\cite{2pszs} also use normal decomposition, but applied to a different
antisymmetric matrix to~\cite{BWG23} (in probability space rather than
logit space, respectively).
\cite{BWG23} explore empirically these different decomposition approaches
as rating schemes, along with the original Elo rating scheme.
%
%\cite{2pszs} visualize their decomposition as $2$-dim embeddings but do 
%not provide theoretical insights to interpret the decomposition.

\begin{figure}[ht]
\newcommand{\fsize}{0.9}
  \centering
  \includegraphics[width=0.49\columnwidth]{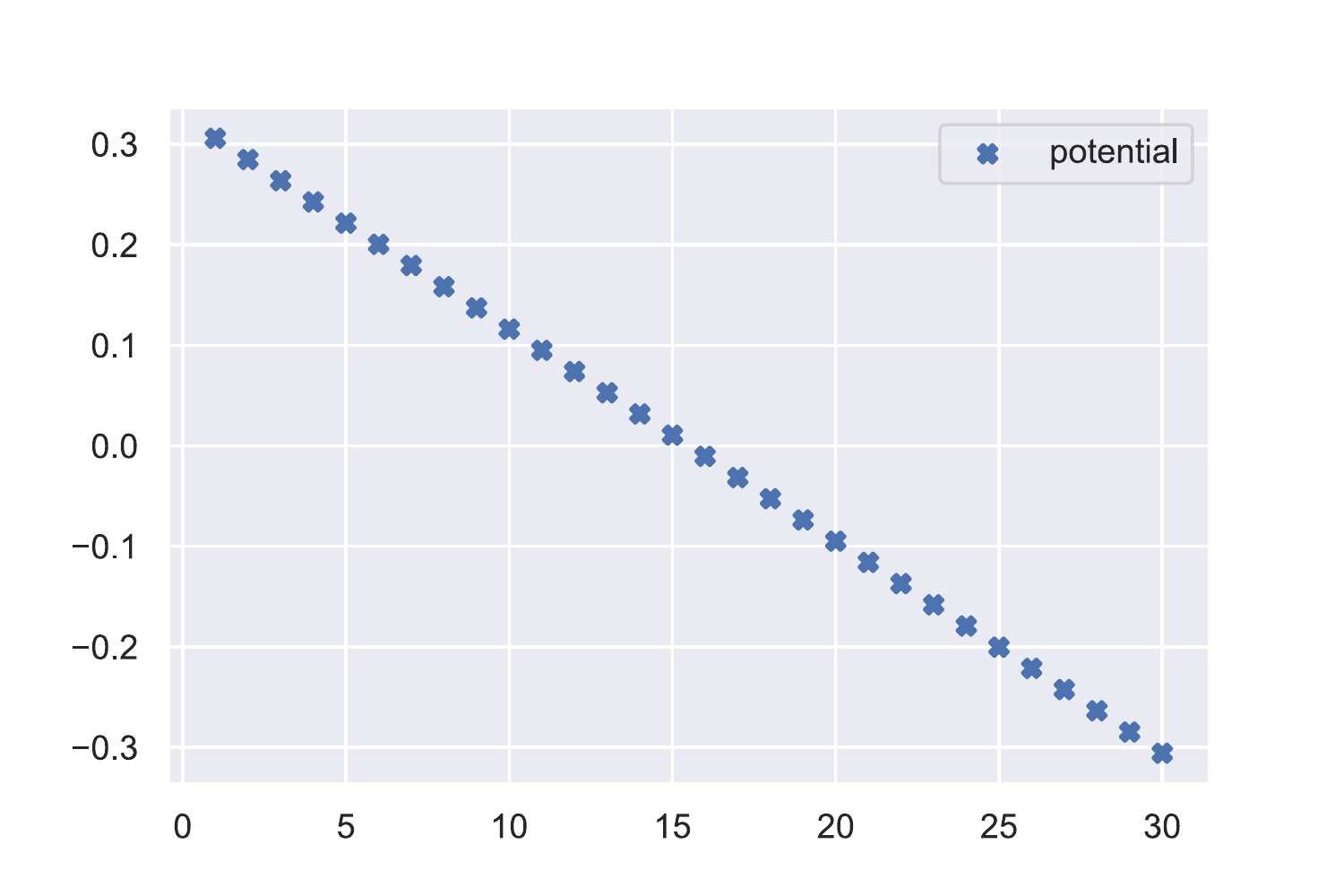}
\includegraphics[width=0.49\columnwidth]{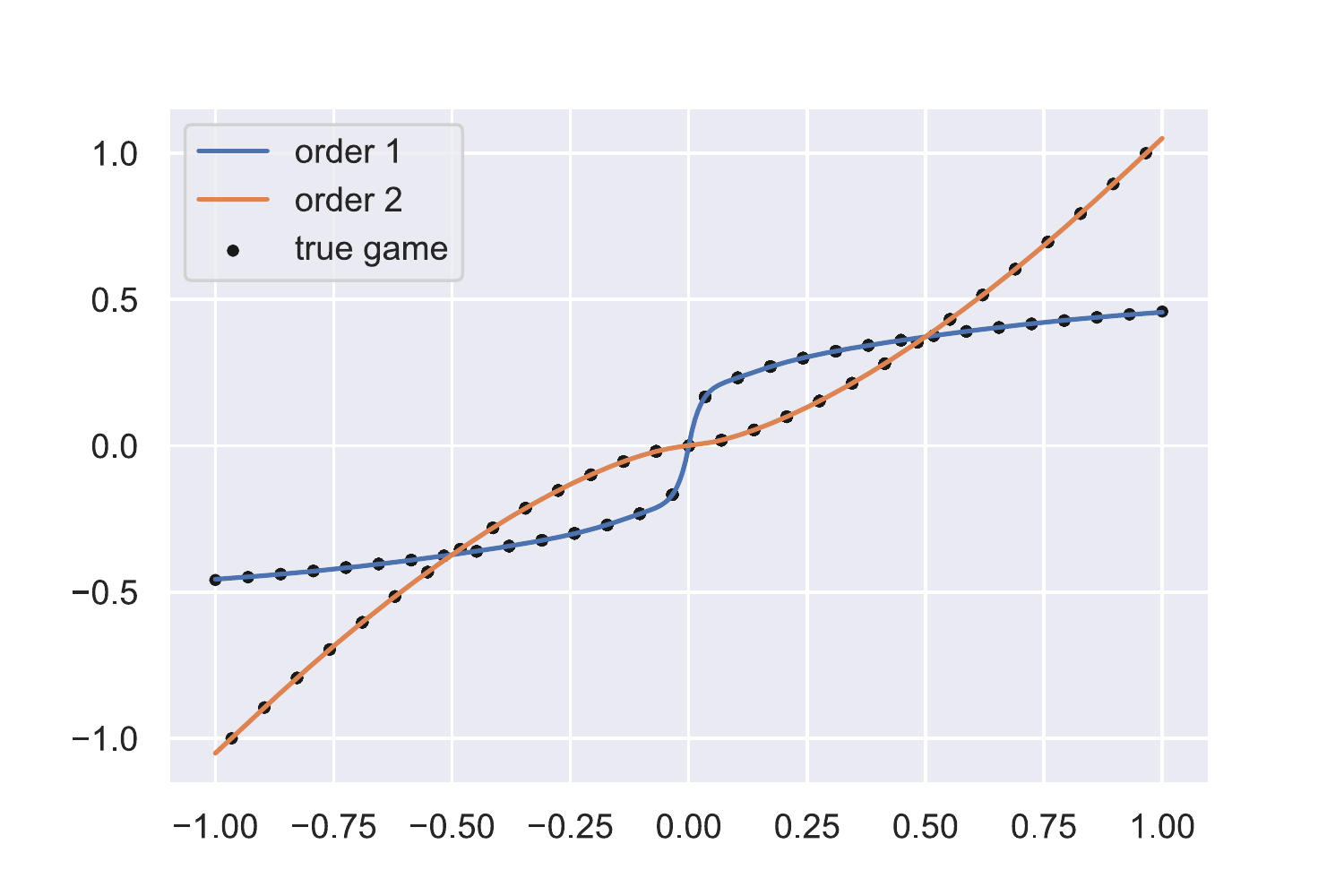}
  \caption{Transitive game of order two of polynomial type, $n=30$. (left) $Y$-axis: ordinal potential player scores $\Phi_i$; $X$-axis: player index; (right) game $P$ and its learnt basis functions as a function of its disk decomposition, cf. Section~\ref{seclearn}: $X$-axis: disk space $\DD$,  $Y$-axis: payoff space $P$. We are able to learn the two generating polynomial functions and that the player scores $\Phi_i$ are evenly spaced (details are provided in Appendix~\ref{app:fig1}).}
  \label{fig_t2}
\end{figure}

\textbf{Outline of the paper.} Our starting point is a result in \cite{BWG23} who observed that Elo ratings do not preserve transitivity, namely that transitive relations among strategies in the original game and its associated Elo game can be different. We show how Elo ratings can be made to preserve transitivity in a simple way by computing these ratings in the right space. We call this approach \textit{hyperbolic Elo rating}: we first transform the game using the invertible mapping $\varphi_{\beta}(x):= \frac{1}{\beta} \tanh(\beta x)$, then compute the Elo ratings, and then go back to the original space using $\varphi_{\beta}^{-1}$. The core idea of the paper is the use of suitable invertible mappings such as $\frac{1}{\beta} \tanh(\beta x)$, that we will call \textit{basis functions} and that we will learn with a neural network. We observed that the approach we used for hyperbolic Elo ratings can actually be extended in much more generality to compute game decompositions of arbitrary games. For this, we transform the original game using possibly multiple basis functions, then compute a game decomposition of the transformed game, and eventually go back to the original space using basis function inverses. The reason why we can do this is that applying basis functions to (entries of) the game does not modify transitive and cyclic relations among strategies. Hence, if one is interested in encoding as efficiently as possible these cyclic and transitive relations, one is free to search for the best basis functions to apply to the game such that the transformed game is as easy as possible to decompose. We show that this amounts to computing the sign-rank of the game, i.e. the minimum rank achievable by a matrix having the same sign pattern\footnote{Sign-rank is important in the theoretical field of communication complexity, where it is studied for arbitrary matrices~\cite{AlonMY16,signrankac0}; here we consider the case of antisymmetric matrices.}. We show that transitive games have sign-rank two, and the number of components needed in our decomposition is essentially the sign-rank. We define the sign-order of a game as the minimum number of basis functions needed to transform the game into a matrix achieving its sign-rank. Elo games are an example of transitive games of order one, and the order can be seen as one measure of the complexity of a hybrid game. The game components in existing methods are in charge of explaining both the sign and amplitude of the payoff. Our neural-network-based approach decouples the learning of the two, which allows us to get important results such as a transitive game always being decomposed using one transitive component that shares the same transitive relations as the game; and a cyclic game being always decomposed using only cyclic components that together share the same cyclic relations as the game. This is not the case in existing methods where for example, a transitive game can be decomposed using only cyclic components. We illustrate on a simple toy example in Figure~\ref{fig_t2} how our method is able to learn a transitive game of order two generated by two polynomials, where player ratings~$\Phi_i$ are evenly spaced.

\textbf{Our contributions.} \emph{Hyperbolic Elo rating.} We introduce a variant of the Elo rating that is guaranteed to preserve transitivity of the original game. \emph{Characterization of transitive games as potential games.} We provide a new characterization of a transitive symmetric zero-sum game as a (weaker) variant of an ordinal potential game with an additively separable potential function. \emph{Decoupled learning approach.} We present a neural-network-based approach that learns a game decomposition into one transitive component and possibly many cyclic components. Contrary to exisiting methods, it decouples the learning of the sign pattern from learning a secondary set of sign-preserving invertible mappings (basis functions) to reconstruct the amplitude of payoff entries.  Our decomposition satisfies that the transitive (resp. cyclic) component of a cyclic (resp. transitive) game is zero. \emph{Empirical evaluation.} We provide a comprehensive evaluation of our methodology using both toy examples and empirical data taken from real-world games. We compare our method to a range of prior approaches \cite{elo1,Sis10,BalduzziTPG18,BWG23,2pszs}
both for complete games and games with missing entries.

%%%%%%%%%%%%%%%%%%%%%%%%%%%%%%%%%%%%%%%%%%%%%%%%%%%%%%%%%%%%%%%%%%%%%%%%%%%%%%%
% FROM BWG23
%%%%%%%%%%%%%%%%%%%%%%%%%%%%%%%%%%%%%%%%%%%%%%%%%%%%%%%%%%%%%%%%%%%%%%%%%%%%%%%
% For ourselves we leverage the symmetric zero-sum game structure, 
% On the other hand, previous works from the pairwise comparison
%community \citep{Shah2017} relied on threshold singular value decomposition
%\citep{Chatterjee2015} to provide statistical guarantees on the payoff matrix
%estimation.

%%%%%%%%%%%%%%%%%%%%%%%%%%%%%%%%%%%%%%%%%%%%%%%%%%%%%%%%%%%%%%%%%%%%%%%%%%%%%%%
% FROM BWG23
%%%%%%%%%%%%%%%%%%%%%%%%%%%%%%%%%%%%%%%%%%%%%%%%%%%%%%%%%%%%%%%%%%%%%%%%%%%%%%%
%While we focus on player strength evaluations for better matchup predictions, a
%related line of work consists of "only" ranking players without matchup
%predictions. \cite{Czarnecki2020} proposed to compute the Nash equilibria of
%empirical games to cluster players into level sets of ``strength''. However,
%when the game has more than two players or is not zero-sum, the computation of
%the Nash equilibrium is in a class of problem complexity called
%PPAD-complete~\citep{chen2009settling,daskalakis2009complexity} which is
%considered intractable. Motivated by this intractability
%result~\cite{omidshafiei2019alpha,rashid2021estimating} proposed a tractable
%ranking technique ($\alpha$-rank) theoretically grounded in the dynamical
%system theory.

%%%%%%%%%%%%%%%%%%%%%%%%%%%%%%%%%%%%%%%%%%%%%%%%%%%%%%%%%%%%%%%%%%%%%%%%%%%%%%%
%%%%%%%%%%%%%%%%%%%%%%%%%%%%%%%%%%%%%%%%%%%%%%%%%%%%%%%%%%%%%%%%%%%%%%%%%%%%%%%

\section{ORDINAL POTENTIAL-BASED PLAYER RATING: FROM ELO TO POTENTIAL GAMES}
\label{secop}

\textbf{Notations.} For any function $\varphi: \R \to \R$ and matrix $A$, we write $\varphi(A)$ for the matrix with entries $\varphi(A_{ij})$. $A^T$ is the transpose of $A$. $\sigma(x):=(1+e^{-x})^{-1}$ is the sigmoid function, and its inverse is the logit function $\logit(x):=\ln(\frac{x}{1-x})=2\arctanh(2x-1)$, so that $2\sigma(x)-1=\tanh(\frac{x}{2})$. We write $\sign(A)$ for the matrix that contains the elementwise sign of $A$, where "sign" can be either $\pm 1$ or $0$. $\bm{1}$ is the vector of all ones.

\textbf{Setup and definitions.} We define a game among~$n$ players via a matrix $\widetilde{P}$ of size $n \times n$ with entries in $[0,1]$ and satisfying $\widetilde{P}_{ij}=1-\widetilde{P}_{ji}$ $\forall i,j$. Following \cite{BWG23}, $\widetilde{P}_{ij}$ can be interpreted as the probability that player $i$ wins against player $j$, i.e. that "$i$ beats $j$". We will sometimes write $i \to j$. We say that there is a tie between $i$ and $j$ when $\widetilde{P}_{ij}=\frac{1}{2}$. Let $P_{ij}:=2\widetilde{P}_{ij}-1$. The matrix $P$ takes value in $[-1,1]$ and is antisymmetric, namely $P=-P^T$. Then "$i$ beats $j$", "$j$ beats $i$" and "$i$ ties with $j$" correspond to $P_{ij}>0$, $P_{ij}<0$ and $P_{ij}=0$, respectively. $P$ is called a (win-loss) payoff matrix in \cite{CzarneckiGTTOBJ20}. We will refer to the game either by $\widetilde{P}$ or $P$. In fact, the matrix $\varphi(P)$ is antisymmetric for any odd function $\varphi$, so one can equivalently see the game $\widetilde{P}$ via $\varphi(P)$, provided that $\varphi$ is positive on $(0,+\infty)$, which preserves the sign of $P$. Common choices \cite{BWG23} are the "probability transform": $\varphi=\id$, and the "logit transform": $\varphi=2\arctanh$, which yields $\varphi(P)=\logit(\widetilde{P})$. The matrix $\varphi(P)$ can be seen as a two-player $n \times n$ symmetric zero-sum "meta game", where each
pure strategy of the meta game corresponds to one of the original~$n$ players~\cite{2pszs} \footnote{Every finite two-player symmetric zero-sum game corresponds to an antisymmetric matrix.}. We now recall the definition of transitive and cyclic games.

\begin{definition}\textbf{(transitivity, cyclicity \cite{BWG23})}
	A game $P$ is transitive if $P_{ij}>0$ and $P_{jk}>0$ implies $P_{ik}>0$ $\forall i,j,k$. $P$ is cyclic if there exists a permutation $\gamma$ of $[1,n]$ such that $P_{\gamma(i)\gamma(i+1)}>0$ $\forall i \in [1,n-1]$ and $P_{\gamma(n)\gamma(1)}>0$. We call a game \emph{hybrid} when it is neither cyclic nor transitive.
\end{definition}

If $(i_1,i_2,...,i_R)$ is a set of indexes, we call $i_1 \to i_2 \to ... \to i_R \to i_1$ a cycle of length $R$. A \emph{maximal cycle} is a cycle with length no less than that of any other cycle. Hybrid games have maximal cycles of length strictly less than $n$, cyclic games have 
		      %at least one 
a maximal cycle of length~$n$. Note that cyclic games are called "fully cyclic" in \cite{BWG23}, whereas \cite{2pszs} uses cyclic for a game with $P\bm{1}=0$. Similarly, the literature has introduced variants in the definition of transitivity, for example \cite{JiangLYY11} allows non-strict inequalities in their definition of "triangular transitivity". Transitive games are also called "monotonic" in \cite{2pszs,CzarneckiGTTOBJ20}. 

We say that two games $P$ and $Q$ have the same sign pattern (in short, sign) and write $P \sim Q$, if $P_{ij}>0\Leftrightarrow Q_{ij}>0$ $\forall i,j$. Note that for any game $P$, $P \sim Q$ implies both $P_{ij}<0\Leftrightarrow Q_{ij}<0$ and $P_{ij}=0\Leftrightarrow Q_{ij}=0$ $\forall i,j$ because the matrices are antisymmetric\footnote{To see why, assume that $P_{ij}=0$ and $Q_{ij}<0$. Then, $P_{ji}=0$ and $Q_{ji}>0$, which is not possible.}. $P$ is said to be regular if there are no ties, namely $P_{ij} \neq 0$ for $i \neq j$. We sometimes choose to work with regular games for clarity of presentation. We comment on this technical aspect in Remark~\ref{remarkreg} in Appendix~\ref{app:B}.

We now recall the definition of "Elo games", named as such because they are generated by Elo ratings. Note that every Elo game is transitive since players' ability to win is measured by a single score.
\begin{definition}\textbf{(Elo game) \cite{2pszs,BWG23}}
\label{elodef}
We write $\elo(P)$ for the game with entries $\elo(P)_{ij}:=2\sigma(\varepsilon^P_i-\varepsilon^P_j)-1$, where $\varepsilon^P:=(\varepsilon^P_i)_{i \in [1,n]}$ is the Elo rating of $P$ which solves the minimization problem \cite{BalduzziTPG18}:
\begin{align}
\label{eloequation}
\begin{split}
&\min_{\varepsilon} \sum_{i,j} \bce \left(\widetilde{P}_{ij}, \sigma(\varepsilon_i-\varepsilon_j)\right),\\ 
&\bce(y,\hat{y}) := - y \ln \hat{y} -(1-y) \ln(1-\hat{y}). 
\end{split}
\end{align}
A game $Q$ is said to be Elo if there exists $P$ such that $Q=\elo(P)$.
\end{definition}

\cite{2pszs,BWG23} have studied game decompositions in terms of "disk components" in Definition~\ref{disk}, and their "normal decomposition" (\ref{normaldecomp}). \cite{BWG23} finds the first $K$ disks so as to minimize a distance to $P$ (or to $2\arctanh(P)$), whereas the $m$-Elo rating of \cite{BalduzziTPG18} does the same after having subtracted the column averages from the original game.
 
\begin{definition}
\label{disk} \textbf{(Disk component) \cite{2pszs,BWG23}}
Given $u$ and $v$ two vectors of size $n$, we write $Disk(u,v):=u v^T - v u^T$ for the antisymmetric matrix of size $n \times n$. 
\end{definition}

Note that the rank of $Disk(u,v)$ is zero if $u= \lambda v$, and two otherwise. If $P$ is an antisymmetric matrix, its normal decomposition states that \cite{greub}:
\begin{align}
\label{normaldecomp}
P = \sum_{k=1}^{K} Disk(u^k,v^k),
\end{align}
where $(u^k,v^k)_{k \in [1,K]}$ is an orthogonal family and $K \leq \lfloor \frac{n}{2} \rfloor$. This is presented in \cite{BalduzziTPG18} as the Schur decomposition of antisymmetric matrices. Antisymmetric matrices always have even rank as their nonzero eigenvalues come in complex conjugate pairs. In \cite{BWG23} it is shown that a disk is either transitive or cyclic, and that a transitive disk can always be written with one of the two vectors having strictly positive entries. This implies that at most one component can be transitive, due to vector orthogonality. One of their motivations to study such decompositions is their observation that the Elo rating fails at preserving transitive relations among players $(i,j,k)$. Namely, if $P$ is transitive, then $P$ and $\elo(P)$ may not necessarily have the same transitive relations among players $(i,j,k)$. We call that succinctly (not) "preserving transitivity". In order to approximate a transitive game, their idea is to consider the transitive disk component, and they show that the latter is able to correctly preserve transitive relations in some examples of transitive games. We show in Proposition~\ref{counterex} that, unfortunately, this is not the case in general, with the proof via a counterexample, which can be found in Appendix~\ref{app:proofprop1}. 
% for which we show that the Hyperbolic Elo rating in theorem~\ref{theoremtanh} allows to preserve transitivity of the original game. 
%
Moreover, we also provide in Appendix~\ref{app:proofprop1} an example that shows that there are (rare) cases when the normal decomposition of a transitive game consists of cyclic components only. The essence of these examples is that nothing forces the components of the normal decomposition to preserve the sign of $P$, whereas transitive and cyclic relations among players depend on the sign only.

\begin{proposition} 
\label{counterex}
\textbf{(The normal decomposition and $m$-Elo do not preserve transitivity)} Let $P$ be a transitive game, and let $\T:=Disk(u^\T, v^\T)$ be the transitive component of the normal decomposition of~$P$. Then, we can have $\sign(\T) \neq \sign(P)$. Similarly, the transitive component of $m$-Elo does not preserve transitivity of $P$. Further, there exists a transitive game $P$ such that its normal decomposition consists exactly of two cyclic components.
\end{proposition}

This motivates us to understand under which conditions we can preserve transitivity. We observe that we do not modify transitive and cyclic relations in a game by applying to each entry an odd function that is positive on $(0,+\infty)$, for example $\varphi_{\beta}(x):= \frac{1}{\beta} \tanh(\beta x)$. We use this idea in Theorem~\ref{theoremtanh} to first transform the game, then compute the Elo rating, then go back to the original space. This shows that it is crucial to compute the Elo rating in the right space if we want to preserve transitivity of $P$.

\begin{theorem}
\label{theoremtanh}\textbf{(Hyperbolic Elo rating preserves transitivity)}
Let the game $P$ be regular and transitive, $0<\alpha<\frac{2}{n(n-1)}$ and $x_\alpha$ be the unique positive root of $2\arctanh^3\left(x\right)-3\alpha x$. Then $P \sim \elo(P)$ provided:
\begin{align*}
 &P_{max}< \frac{x_\alpha}{n-1} \hspace{4mm} \mbox{ and } \hspace{4mm}\frac{P_{max}}{P_{min}} < \frac{n}{n-2 + n (n-1) \alpha},
\end{align*}
where $P_{max}:= \max_{i,j} P_{ij}$, $P_{min}:= \min_{P_{ij}>0} P_{ij}$. In particular, let $\varphi_{\beta}(x):= \frac{1}{\beta} \tanh(\beta x)$ and $\beta_\alpha>0$ such that:
\begin{align*}
\frac{\tanh(\beta_\alpha P_{max})}{\tanh(\beta_\alpha P_{min})} \leq \frac{n}{n-2 + n (n-1) \alpha},
\end{align*}
for example $\beta_\alpha = \frac{1}{P_{min}}\arctanh \left( \frac{n-2}{n} + (n-1) \alpha\right)$ \footnote{Due to $0<\alpha<\frac{2}{n(n-1)}$.}. Then $P \sim \varphi_{\beta}(P) \sim \elo(\varphi_{\beta}(P))$ provided $\beta > \max \left( \frac{n-1}{x_\alpha}, \beta_\alpha\right)$. That is, the rating system $\widehat{P}:=\varphi_{\beta}^{-1}(\elo(\varphi_{\beta}(P)))$ preserves transitivity for high enough~$\beta$, namely $P \sim \widehat{P}$ \footnote{We have $\varphi_{\beta}^{-1}(x) = \frac{1}{\beta} \arctanh(\beta x)$ and we use the convention that $\varphi_{\beta}^{-1}(x):=1$ for $x \geq \varphi_{\beta}(1)$, $\varphi_{\beta}^{-1}(x):=-1$ for $x \leq -\varphi_{\beta}(1)$, so that $\widehat{P}$ takes value in $[-1,1]$.}.
\end{theorem}

Theorem~\ref{theoremtanh} essentially states that the Elo rating preserves transitivity if the gap between $P_{max}$ and $P_{min}$ is not too big. This yields a straightforward recipe to guarantee that transitivity is preserved, which we call Hyperbolic Elo rating: first compute $\varphi_{\beta}(P)$ for high enough $\beta$, then compute the Elo rating of $\varphi_{\beta}(P)$, then go back to the original space by applying $\varphi_{\beta}^{-1}$. The main merit of the formulas presented in Theorem~\ref{theoremtanh} is that they are explicit. In practice, it is possible to get tighter bounds, which we discuss in Remark~\ref{remarkbeta} in Appendix~\ref{app:B}, together with the case where the game is not regular, in which case we still get $P_{ij}>0 \Rightarrow \elo(P)_{ij}>0$.

It is known that an Elo game is transitive, but the converse is false \cite{BWG23}. Therefore, a suitable characterization of transitive games seems to be lacking in the literature. It is of interest to ask \textit{what is a transitive game?} We provide two such characterizations. The first one in Theorem~\ref{theorempotential} links transitive games to potential games, a fundamental concept in game theory. The second one in Corollary~\ref{corsignrank} reformulates Theorem~\ref{theorempotential} using the concept of sign-rank.

We recall from the seminal paper that introduced potential games~\cite{potential} that a two-player symmetric zero-sum game defined via the antisymmetric matrix $P$ is an \textit{ordinal potential} game if there exists a matrix $\widetilde{\Phi}$ such that $\forall i,j,k$:
\begin{align}
\label{op}
 P_{ij}-P_{kj} > 0 \Leftrightarrow \widetilde{\Phi}_{ij}-\widetilde{\Phi}_{kj} > 0 \Leftrightarrow \widetilde{\Phi}_{ji}-\widetilde{\Phi}_{jk} > 0.
\end{align}

We call $\widetilde{\Phi}$ a potential function, or more succinctly a potential. In general, a bimatrix game with players' payoffs $A$ and $B$ is an ordinal potential game if $A_{ij}-A_{kj} > 0 \Leftrightarrow \widetilde{\Phi}_{ij}-\widetilde{\Phi}_{kj} > 0$ and $B_{ji}-B_{jk} > 0 \Leftrightarrow \widetilde{\Phi}_{ji}-\widetilde{\Phi}_{jk}>0$. When the game is zero-sum and symmetric, $B=-A=A^T$, so the latter is equivalent to (\ref{op}). Note that $\widetilde{\Phi}$ need not be symmetric, for example one could have $P_{ij}:=\Phi_{i}-\Phi_{j}$, and in that case $\widetilde{\Phi}_{ij} := \alpha(\Phi_{i}) + \beta(\Phi_{j})$ is an ordinal potential for every pair of strictly increasing functions $\alpha, \beta$. This implies in particular that ordinal potentials are not unique in general, contrary to exact potentials (which are unique up to an additive constant \cite{potential}). We first define a weak variant of ordinal potential games that is obtained by taking the special case $j=k$ in the definition of ordinal potential games (\ref{op}).

%It has been shown in \cite{brm03} that a two-player zero-sum game is an exact potential game if and only if it has a separable payoff function. Theorem~\ref{theorempotential} can be seen as a result of the same flavor but for ordinal potential games. Most importantly, it gives a characterization of transitivity.

\begin{definition}
\label{potentialdef} \textbf{(weak separable ordinal potential game)}
A two-player symmetric zero-sum game $P$ is a weak ordinal potential game if (\ref{op}) holds for all $i$ and all $j=k$ \footnote{As opposed to $\forall i,j,k$ for ordinal potential games.}. It is a separable ordinal potential game if $\widetilde{\Phi}$ in (\ref{op}) is additively separable, namely $\widetilde{\Phi}_{ij} = \alpha_{i} + \beta_{j}$. It is a weak separable ordinal potential game if it is both of the above.
\end{definition}

Theorem~\ref{theorempotential} is the main result of this section as it characterizes transitive games. The direction "$\Leftarrow$" is immediate, the other direction is more challenging. It is in fact a consequence of Theorem~\ref{theoremtanh}.

\begin{theorem}
\label{theorempotential} \textbf{(transitive $\Leftrightarrow$ weak separable ordinal potential)}
A regular game $P$ is transitive if and only if it is a weak separable ordinal potential game, namely there exists a vector $\Phi$ such that:
\begin{align*}
 P_{ij} > 0 \Leftrightarrow \Phi_{i}-\Phi_{j} > 0 \hspace{6mm} \forall i,j.
\end{align*}
The potential $\Phi$ can be chosen as the Elo rating $\varepsilon^{\varphi_{\beta}(P)}$ of $\varphi_{\beta}(P)$, where $\varphi_{\beta}$ is as in Theorem~\ref{theoremtanh}.
\end{theorem}

The proofs of Theorems~\ref{theoremtanh} and~\ref{theorempotential} were made with non-regular games in mind, and we comment on this aspect in Appendix~\ref{app:proofthm1}. The proof yields that if a game $P$ is transitive (but not necessarily regular), then it is a weak separable generalized ordinal potential game, namely:
$$
P_{ij} > 0 \Rightarrow \Phi_i - \Phi_j> 0 \hspace{5mm} \forall i,j,
$$
where the term "generalized ordinal potential" has been introduced in \cite{potential} and means that we only have "$\Rightarrow$" instead of the "$\Leftrightarrow$" that we have for ordinal potential games. 

The function $\varphi_{\beta}$ in Theorem~\ref{theoremtanh} was chosen ad hoc. This naturally brings the question of optimality of such functions, in the sense that they allow one to better reconstruct the game from the potential. This leads to our definitions of basis functions and sign-order.
 
\begin{definition}
\label{basisfunction} \textbf{(Basis function)}
A function $\varphi:\R \to \R$ is said to be a basis function if it is odd and strictly increasing.
\end{definition}
We build on our characterization of transitive games to define the new concept of sign-order, which is the minimum number of basis functions needed to move between the payoff and the potential function (and vice versa; basis functions are invertible by definition).

Let $\bm{\varphi}:=(\varphi_m)_{m \in [1,M]}$ be a collection of basis functions. For an antisymmetric matrix $A$, we write $\M_{\bm{\varphi}}(A)$ for the set of matrices for which each entry $(i,j)$ is the image of $A_{ij}$ under some basis function:
\begin{align}
\label{varphiset}
\begin{split}
&\M_{\bm{\varphi}}(A):= \{B:\hspace{1.5mm} \forall (i,j), \hspace{1.5mm} \exists\ m(i,j) \in [1,M] \\
&\mbox{ such that }  \hspace{1.5mm} B_{ij} = \varphi_{m(i,j)}(A_{ij})\}.
\end{split}
\end{align}

\begin{definition}
\label{deftransorder} \textbf{(Sign-order)} The sign-order (in short, "order") of a game $P$ is defined as the minimum number $\tau_P$ of basis functions $\bm{\varphi}:=(\varphi_m)_{m \in [1,\tau_P]}$ such that $P \in \M_{\bm{\varphi}}(A)$ for some antisymmetric $A \sim P$ such that $\rank(A) = \min \{\rank(B): B \sim P, \hspace{3mm} B \mbox{ antisymmetric}\}$. In particular, if $P$ is regular and transitive, Theorem~\ref{theorempotential} yields that $A$ can be chosen to be of the form $A_{ij}:=v^\T_i v^\T_j (\Phi_{i}-\Phi_{j})$, where $v^\T$ has strictly positive entries.
\end{definition}

The last claim in Definition~\ref{deftransorder} follows from the following: a non zero antisymmetric matrix has rank at least 2. If $P$ is regular and transitive, then in particular it is non zero. By Theorem~\ref{theorempotential}, the sign of $P_{ij}$ is that of $\Phi_{i}-\Phi_{j}$, which is also that of $v^\T_i v^\T_j (\Phi_{i}-\Phi_{j})$ as $v^\T$ has strictly positive entries. Finally, the entries of any transitive antisymmetric matrix of rank 2 can be written in the form $v^\T_i v^\T_j (\Phi_{i}-\Phi_{j})$, as shown in~\cite{BWG23} (Proposition 2). 

An Elo game is transitive of order one, with its basis function equal to the sigmoid function $\sigma$. There are many other games that also are of order one, for example "polynomial" games $P_{ij}:= \alpha \sign(\Phi_{i}-\Phi_{j}) \cdot |\Phi_{i}-\Phi_{j}|^m$, where $\alpha>0$ is a normalizing constant so that  $P_{ij} \in [-1,1]$. Transitive games of higher order are thus in some sense further from being an Elo game, and orders can be seen as providing a classification of transitive games which can be used, for example, to generate different classes of such games. Since we know that $P_{ij} > 0 \Leftrightarrow \Phi_{i}-\Phi_{j} > 0$, we can always find basis functions by defining, when $P_{ij}>0$, $\varphi_{ij}(x) := \frac{\Phi_{i}-\Phi_{j}}{P_{ij}} x$, so that $\varphi_{ij}(P_{ij})=\Phi_{i}-\Phi_{j}$. Note that $\varphi_{ji}=\varphi_{ij}$, so there are in the worst case $\frac{n(n-1)}{2}$ unique such functions. Even when the game is transitive of order one, the method we introduce in Section~\ref{seclearn} allows us to learn~$\varphi$ rather than postulating it as in existing work. We illustrate in Appendix~\ref{app:A} and in Figure~\ref{fig_t2} examples of transitive games of polynomial type, of sign-order one and two. The concepts of potentials and sign-order are also useful for arbitrary games.

\begin{figure}[ht]
\newcommand{\fsize}{1.}
  \centering
  \includegraphics[width=\columnwidth]{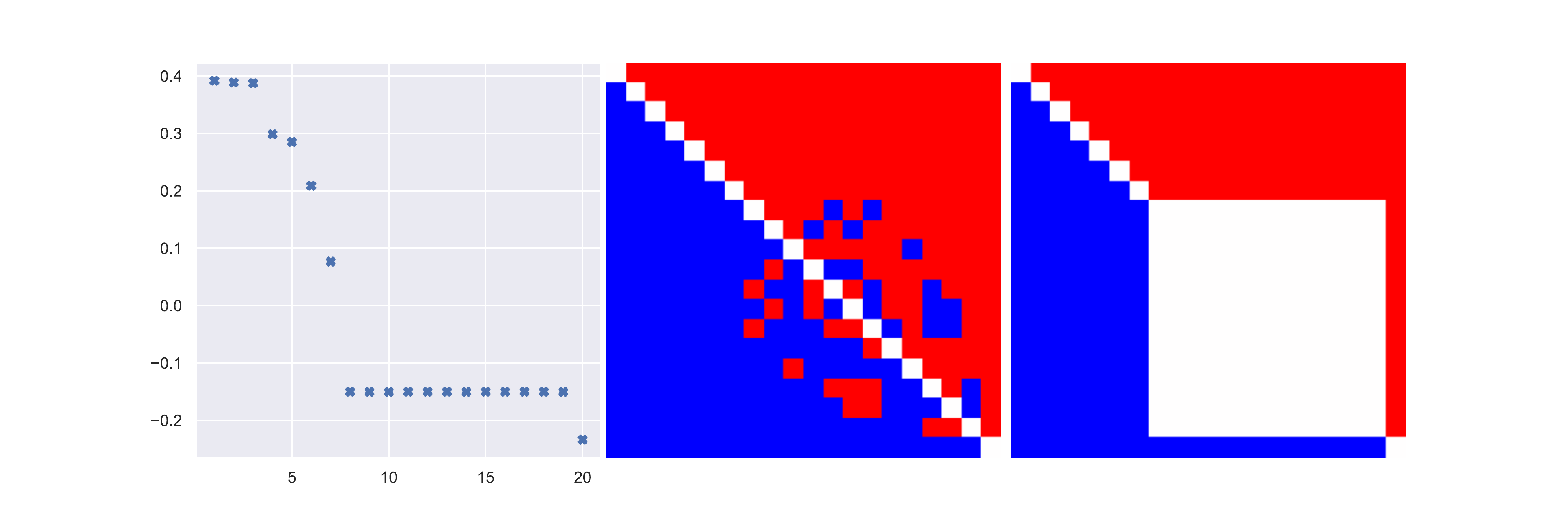}
  \caption{AlphaStar, $n=20$. (Left) ordinal potential-based ratings of the $n$ players in decreasing order; (middle) sign of the game $P$; (right) sign of the potential component $\Phi_i-\Phi_j$. Red is positive, blue is negative, white is zero. We are able to learn correctly the 9 rating "levels": first, 7 players each with their own rating, then a large cycle where players share the same rating, and finally a player who loses against everyone.}
  \label{fig_alphas_tc}
\end{figure}

\textbf{Transitive (ordinal potential) component of an arbitrary game.} From Theorem~\ref{theorempotential}, we define a transitive component of the game $P$ as a matrix $\T=Disk(\Phi \odot v^\T,v^\T)$ such that $\T_{ij} P_{ij} \geq 0$, $v^\T$ has strictly positive entries and $\odot$ is the elementwise product, i.e. $\T_{ij}=v^\T_i v^\T_j (\Phi_{i}-\Phi_{j})$. For example $\T=Disk(\Phi,\bm{1})$. It is immediate that if $2$ players $i,j$ are in the same cycle, then $\Phi_i=\Phi_j$. So, if the game is cyclic, the potential component $\T=0$. In the case of a hybrid game, every player is either in some cycle, or in no cycle. In the former case, we expect our method -- described in the next section -- to learn the same rating for all players in a given cycle. This is illustrated in Figure~\ref{fig_alphas_tc} in the case of AlphaStar data with $n=20$ \footnote{We provide more details on the learning algorithm and data in Section~\ref{secexp} and in the appendix.}. We see that in this case, we first have a set of $7$ players such that $i$ wins against $i+1$. Then, we have a large cycle containing 12 players, and finally we have a player who loses against everyone. We see that we are able to learn correctly the ratings $\Phi$ with the method presented in Section~\ref{seclearn} (we considered one basis function $\varphi$ and provide in the appendix the learnt game and components). This approach is similar to the "layered" geometry in \cite{CzarneckiGTTOBJ20} where transitivity is viewed as the index of a cluster. In our case we also learn scores to assign to each such layer.

\section{LEARNING TO DECOMPOSE AN ARBITRARY GAME}
\label{seclearn}

In this section we describe the methodology that we will use to learn ordinal potential-based ratings. We will actually go further and learn cycles as well.

\textbf{Sign-rank and the learning of cycles.} The \emph{sign-rank} of a matrix $P$ with entries $\pm 1$ is the minimum rank achievable by a real matrix with entries that have the same sign as those of $P$ \cite{AlonMY16,signrankac0}. We will say that a matrix $Q$ achieves the sign-rank of $P$ when $Q$ has the same sign as $P$ and the rank of $Q$ is equal to the sign-rank of $P$. One can see a matrix achieving the sign-rank as the most efficient encoding of the sign of $P$. It is trivial to extend the definition of sign-rank to the case where entries of $P$ can take arbitrary non-zero values, since in that case one can consider $\sign(P)$ to get back to the canonical case. We further extend the definition of sign-rank as follows: we allow entries to take the value zero, so that the sign can take value $\pm 1$ or $0$, and we restrict ourselves to minimum ranks achievable by antisymmetric matrices. This yields Definition~\ref{defsignrank}.

\begin{definition}
\label{defsignrank} \textbf{(Sign-rank of an antisymmetric matrix)}
The sign-rank of an antisymmetric matrix $P$ is the minimum rank achievable by an antisymmetric matrix $Q \sim P$.
\end{definition}

We get in Corollary~\ref{corsignrank} a reformulation of Theorem~\ref{theorempotential} using the concept of sign-rank. 
\begin{corollary}
\label{corsignrank}
A regular game $P$ is transitive if and only if there exists a disk $Disk(u^\T,\bm{1})$ achieving the sign-rank, i.e. $P \sim Disk(u^\T,\bm{1})$. In particular, any regular transitive game has a sign-rank of two, and the vector $u^\T$ can be chosen as the Elo rating of $\varphi_{\beta}(P)$.
\end{corollary}

There exists cyclic games of sign-rank two such as rock-paper-scissors, however in this case neither $u$ nor $v$ can be equal to $\bm{1}$. It turns out that cyclic games can be decomposed using cyclic disks only.  
\begin{theorem}
\label{propcyclic1}
\textbf{(Cyclic games can be decomposed using cyclic disks only)}
A regular game $P$ is cyclic with a maximal cycle $\OO$ if and only if $P \sim \sum_{k=1}^{K} Disk(u^k,v^k)$ for some $K$ and some vectors $u^k$, $v^k$, where each disk $Disk(u^k,v^k)$ is cyclic and admits $\OO$ as a maximal cycle.
\end{theorem}

Consequently, the cyclic component in our decomposition will consist of cyclic disks only. We provide in the appendix an upper bound on the number of cyclic disks that one can expect in Theorem~\ref{propcyclic1}. Counting the minimal number of cyclic disks required to capture the sign of $P$ is challenging due to the compensation effect between these disks. It is tempting to believe that in the case of a cyclic game, one can achieve the sign-rank using only cyclic disks. This is what we are able to do in Figure~\ref{fig_cyclic_2} where we learn the correct sign with three disks, all cyclic as in Theorem~\ref{propcyclic1}. The normal decomposition is not able to learn the correct sign with 3 disks, and furthermore one of the learnt disks is transitive, which is counterintuitive for a cyclic game. We provide Conjecture~\ref{conj} that we leave for future work.

\begin{conjecture}
\label{conj}
A regular game is cyclic if and only if its sign-rank is achievable by cyclic disks only.
\end{conjecture}

Let $\T:=Disk(u^{\T},v^{\T})$ be the transitive component, $\C:=\sum_{k=1}^{K} Disk(u^k,v^k)$ the cyclic component, and let $\DD:=\T + \C$ be our decomposition. We call the latter the \textit{\textbf{disk space}}, which can be seen as a "latent space". Definition~\ref{deftransorder} yields that the order is the minimum number of basis functions needed to move from the game $P$ to its disk representation $\DD$. We will require $v^{\T}$ to have strictly positive entries, so that $\T$ is by construction transitive. We will also require $u^k$, $v^k$ to be orthogonal to each other and to $v^{\T}$, so that $\C$ will consist by construction of cyclic components (cf. discussion below Definition~\ref{disk}). Orthogonality ensures, in short, that there is no redundancy between components in the "linear algebra" sense. In practice, we have seen that it makes the learning faster. 

The \textit{number} of components in our decomposition aims at correctly capturing the sign of $P$. Precisely, for a budget of $K$ components, we aim to minimize the number of entries $\LL_{sign}(\DD, P)$ which have different signs in $\DD$ and $P$. Let $\Ss_{P}(K)$ be that number, and $\Ss_{P} := \min\{K: \Ss_{P}(K)=0\}$. $\Ss_{P}(K)$ quantifies the ability of $K$ components to capture $\sign(P)$. The normal decomposition ensures that there exists a $K$ such that $\Ss_{P}(K)=0$, and the sign-rank of $P$ is equal to $2\Ss_{P}+2$ or to $2\Ss_{P}$\footnote{The sign-rank of $P$ is $2\Ss_{P}+2$ if $P$ is transitive ($\Ss_{P}=0$) or hybrid. If Conjecture~\ref{conj} is true, the sign-rank of $P$ is $2\Ss_{P}$ if $P$ is cyclic; if it is false it could be $2\Ss_{P}$ or $2\Ss_{P}+2$.}.

\begin{figure}[ht]
\newcommand{\fsize}{0.49}
  \centering
  \includegraphics[width=\fsize\linewidth]{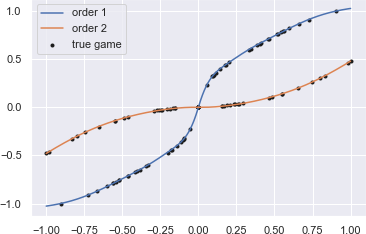}
  \includegraphics[width=\fsize\linewidth]{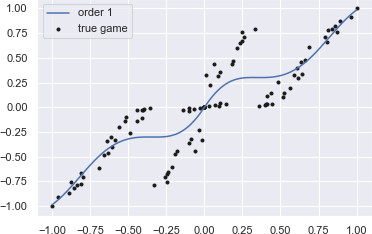}
  \includegraphics[width=\fsize\linewidth]{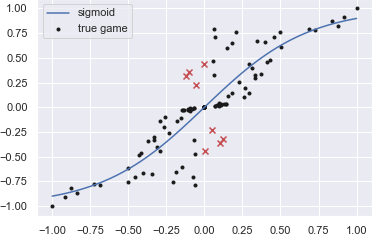}
  \caption{Cyclic game of sign-rank $2$ and order $M=2$. (Top left) ours, 2 basis functions; (top right) ours, 1 basis function; (bottom) normal decomposition. Black dots represent the true game $P$; red crosses indicate the true game when there is a mistake on the sign. All methods learn $K=1$ disk component. Contrary to the baseline, our method is able to learn the sign of $P$, as well as the two functions generating the game. $X$-axis: disk space $\DD$,  $Y$-axis: payoff space $P$.}
  \label{fig_cyclic_1}
\end{figure}

\begin{figure}[ht]
  \centering
  \includegraphics[width=\columnwidth]{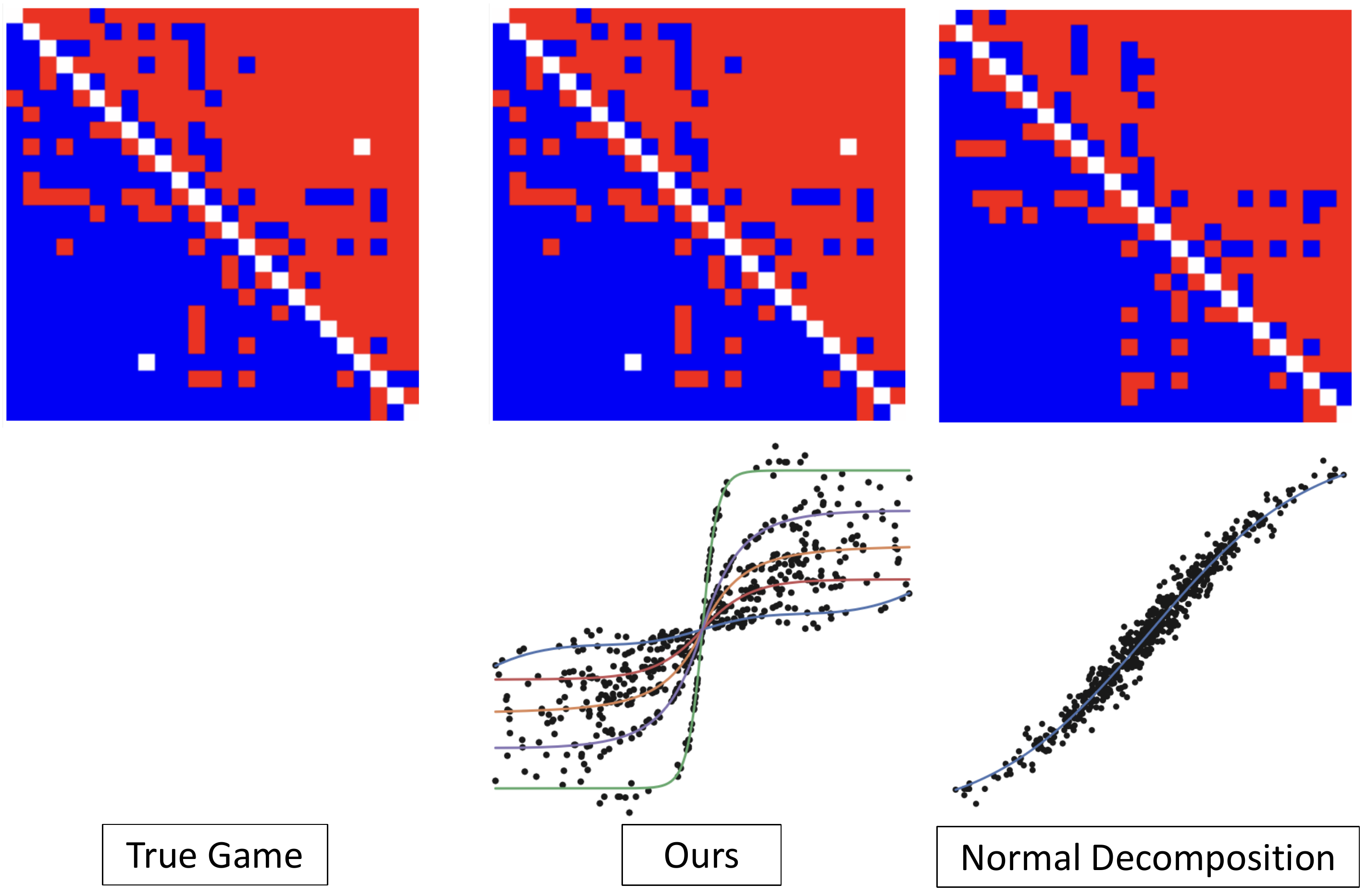}
  \caption{Kuhn-poker, $n=25$ (cyclic game). (Left) true game; (mid) ours; (right) normal decomposition; (top) sign of the game; (bottom) the game in disk space ($X$-axis: disk space $\DD$,  $Y$-axis: payoff space $P$). We are able to learn the sign of the game with 3 components, all cyclic as in Theorem~\ref{propcyclic1}. The normal decomposition cannot learn the correct sign with 3 components, and further one component is transitive, which is counterintuitive for a cyclic game. Red is positive, blue is negative, white is zero.
  }
  \label{fig_cyclic_2}
\end{figure}

Given $M$ basis functions $\bm{\varphi}:=(\varphi_m)_{m \in [1,M]}$, and $K$ disk components, we try to minimize $\LL_{proba}(A, P)$ under the constraint $\LL_{sign}(\DD, P) = \Ss_{P}(K)$, where $A$ ranges over $\M_{\bm{\varphi}}(\DD)$ and $\LL_{proba}$ is a distance on the space of matrices, for example the $L_2$ distance or the binary cross-entropy. In simple words, under the constraint that we do as well as possible on the sign with $K$ components, we play on $\DD$ and $\bm{\varphi}$ to reconstruct $P$ as well as possible. Basis functions do not change the sign of a matrix and the order $\tau_P$ is equal to the minimum $M$ that yields $\LL_{proba}(A, P)=0$. We illustrate these concepts in Figure~\ref{fig_cyclic_1} in the context of a cyclic game of order $M=2$ and sign-rank $2K=2$. The exact definition of the game is provided in the appendix.

\begin{figure}[ht]
  \centering
  \includegraphics[width=\columnwidth]{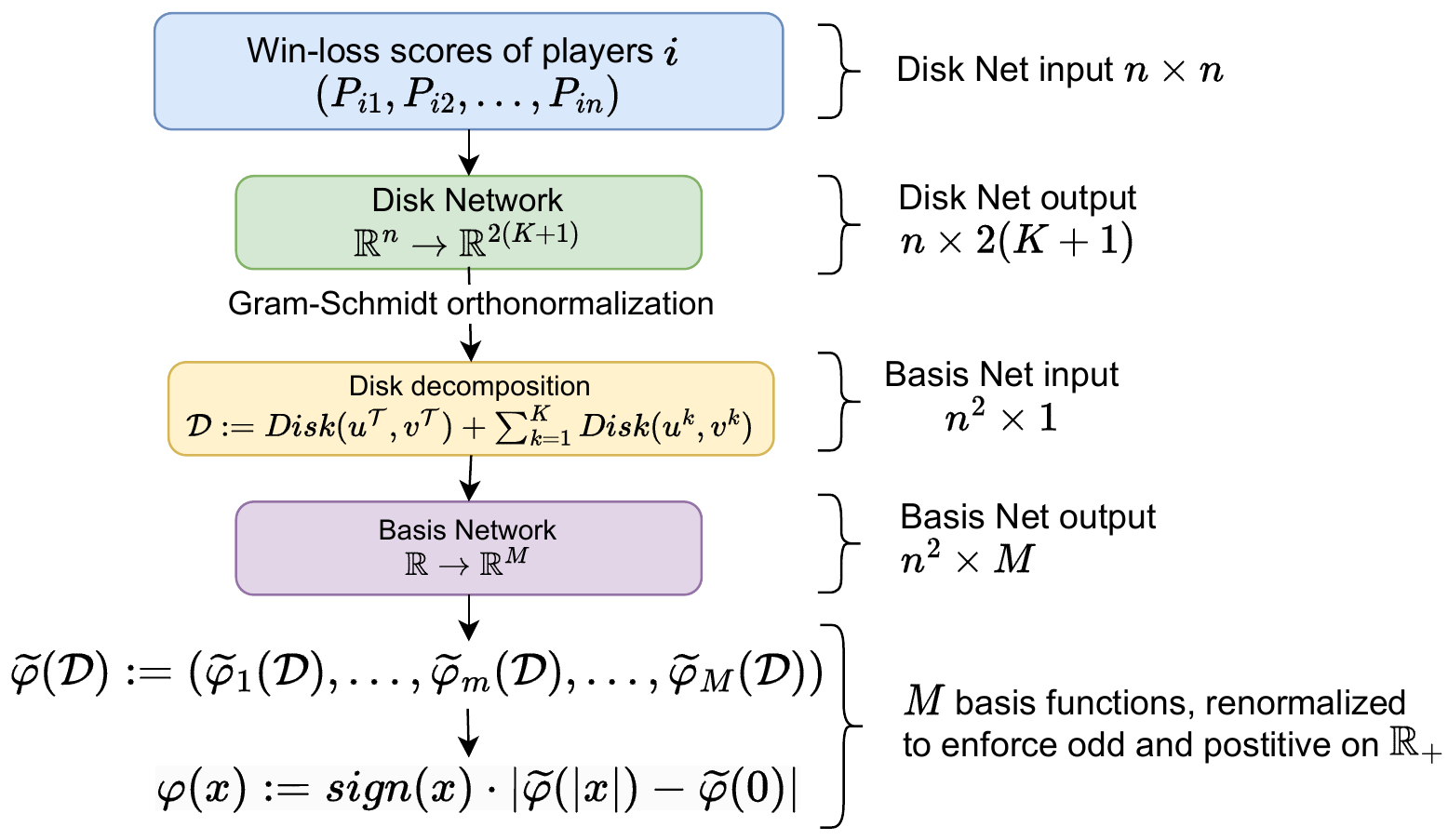}
  \caption{Neural Architecture to learn our game decomposition.}
  \label{fig_neural}
\end{figure}

\textbf{Description of the network architecture.} We provide in Figure~\ref{fig_neural} an overview of our architecture. We first feed the $n \times n$ matrix $P$ into the disk network $\R^n \to \R^{2K+2}$ which outputs the $2(K+1)$ entries $u^k$, $v^k$, $u^\T$, $v^\T$ of our disk decomposition $\DD$, for each player $i \in [1,n]$. Then, we construct $\DD$, and guarantee orthogonality of the vectors by performing Gram-Schmidt orthogonalization in the computational graph. Then, the $n^2 \times 1$ disk-space decomposition $\DD$ is fed into the basis network, which outputs the quantities $\varphi_m(\DD_{ij})$ for $m \in [1,M]$ that we use to compute a reconstruction $\widehat{P}_{ij}$ of $P_{ij}$. At training time, for each matrix entry $(i,j)$, we pick the index $m(i,j)$ that yields the reconstruction $\varphi_{m(i,j)}(\DD_{ij})$ closest to $P_{ij}$. At test time, given a point in the disk-space $\DD_{ij}$, we compute weights $\omega_m(\DD_{ij})$ from the point closest to $\DD_{ij}$ in the training set, and use those weights to compute a prediction $\widehat{P}_{ij}= \sum_{m=1}^M \omega_m(\DD_{ij}) \varphi_m(\DD_{ij})$. The weight $\omega_m$ represents the proportion of training points that were associated to the basis function $m$ at training time. Notice that we suitably transform the functions $\varphi_m$ to make them basis functions, cf. $\varphi$ and $\widetilde{\varphi}$ in Figure~\ref{fig_neural}.

\textbf{Loss function.} Our loss function $\mathcal{L}$ consists of four terms.
\begin{align}
\label{lossf}
\mathcal{L} = \mathcal{L}_{proba} + \omega^{\T}_{sign} \mathcal{L}^{\T}_{sign} + \omega^{\C}_{sign}\mathcal{L}^{\C}_{sign} + \omega_{basis}\mathcal{L}_{basis}.
\end{align}
The term $\mathcal{L}_{proba}$ is simply the reconstruction loss on $P$ discussed earlier; we take the standard mean-squared loss but we could also consider the binary cross-entropy. As previously discussed with $\Ss_{P}(K)$, we put emphasis on learning the sign of the game. Due to Theorem~\ref{theorempotential}, the sign of $P_{ij}$ should either be captured by $\C_{ij}$ if $i$ and $j$ are in the same cycle, otherwise by $\T_{ij}$. Therefore, $\mathcal{L}^{\T}_{sign}$ ensures that $\T$ and $P$ have the same sign in a weak sense as discussed at the end of Section~\ref{secop}: $\T_{ij} P_{ij} \geq 0$ and $P_{ij}=0 \Rightarrow \T_{ij}=0$. This can always be achieved whether $i$ and $j$ are in the same cycle ($\T_{ij}=0$) or not ($\T_{ij} \neq 0$) and therefore we typically pick $\omega^{\T}_{sign}$ very large. Similarly, $\mathcal{L}^{\C}_{sign}$ ensures that $\C_{ij} P_{ij} \geq 0$ and that $P_{ij}=0 \Rightarrow \C_{ij}=0$. If $i$ and~$j$ are in the same cycle, we have $\T_{ij}=0$, and ${\C_{ij} P_{ij} >0}$ cannot always be achieved as this depends on the budget $K$, so for this reason we typically pick $\omega^{\C}_{sign}<\omega^{\T}_{sign}$ but $\omega^{\C}_{sign}>1$ since we want to put emphasis on learning the sign vs. the amplitude $\mathcal{L}_{proba}$. Finally, $\mathcal{L}_{basis}$ aims at ensuring that the basis functions are increasing. We do so by calling the basis network a second time with permuted inputs and considering a loss that penalizes $(X-X^{perm}) \cdot(Y-Y^{perm})<0$. Typically $\omega_{basis}$ is very large since we can always choose the basis functions to be increasing. $\mathcal{L}^{\T}_{sign}$, $\mathcal{L}^{\C}_{sign}$, $\mathcal{L}_{basis}$ are constructed in the spirit of the Pearson correlation coefficient, and are written explicitly in Appendix~\ref{app:loss}, together with the values of $\omega^{\T}_{sign}$, $\omega^{\C}_{sign}$, $\omega_{basis}$; in particular we make sure to suitably normalize them by the norms of $\T$ and~$\DD$, so that learnt coefficients are not pushed towards~0. 

\section{EXPERIMENTS}
\label{secexp}

We consider some of the game payoffs studied in \cite{CzarneckiGTTOBJ20,BWG23} and take the payoff matrices $P$ from the open-sourcing of these works. The baselines that we consider are those in \cite{BWG23}, that is the normal decomposition and $m-$Elo previously discussed. We see in Table~\ref{tabsign} on a variety of games that our method yields better accuracy on the sign of $P$. We report standard deviations as well as other metrics of interest in the appendix. The baselines perform well in general, and are faster to compute than our neural approach. Both our basis and disk networks have 3 hidden layers and 200 neurons per layer. All activation functions are $\tanh$, except for the output of the disk network for which the activation function is the identity. All methods learn $K=3$ components, but additionally we learn the transitive (potential) component. If the game contains a cycle of length $n$ we disable the learning of the potential component for simplicity, which is the case for most of the games in Table~\ref{tabsign} because we only looked at a subset of players. We illustrated the efficacy of our method for learning the potential component on AlphaStar data in Figure~\ref{fig_alphas_tc}. These results, together with those in Figures~\ref{fig_cyclic_1} and~\ref{fig_cyclic_2} show that our method learns more efficiently the sign of the game and hence cyclic and transitive relations among strategies.

\begin{table}[ht]
  \caption{Average sign accuracy (\%) over 3 seeds and game sizes $n=50,75,100$. $K=3$ components.}
  \label{tabsign}
  \centering
  \begin{tabular}{ccccc}
    \toprule
     Game & Elo  & $m$-Elo  & Normal & Ours \\
    \midrule
    connect four & 86 &   94  &  94    &    \bf{97}\\
    5,3-Blotto  &71 &  \bf{99}&   99    &   \bf{99} \\
    tic tac toe &93  &  96  &  96   &     \bf{98} \\
    Kuhn-poker &81  &  91  &  92   &     \bf{96} \\
    AlphaStar &86  &  92 &   92   &    \bf{95}\\
    quoridor(size 4) & 87 &   92  &  93   &      \bf{96}\\
    Blotto & 77  &  94 &   \bf{95}   &      \bf{95} \\
    go(size 4) & 84 &   93 &   93  &    \bf{97} \\
    hex(size 3) &93  &  96 &   97   &   \bf{98}\\
    \bottomrule
  \end{tabular}
\end{table}

\section{CONCLUSION AND FUTURE RESEARCH}
In this work we have characterized the essence of transitivity in games by connecting it to important concepts such as potential games and sign-rank. We have provided a neural network-based architecture to learn game decompositions is a way that puts specific emphasis on the sign of the game. 

We believe that it would be interesting to resolve Conjecture~\ref{conj}, as well as improve the efficiency of the architecture to have it work on larger game sizes. For example, it would be interesting to consider a transformer architecture as in \cite{nfgtransformer}, as the attention mechanism could be useful in learning complex dependencies between the disk representation and the original game.

Another aspect that would be interesting to study is the online update of Hyperbolic Elo and Potential-based ratings. We briefly comment on this aspect below. Assume that at each stage $t+1$, two players are chosen at random to play against each other. As discussed in \cite{elo2,BalduzziTPG18} and assuming that players $i$ and $j$ have been chosen, the online updates of the Elo ratings $\epsilon_i$, $\epsilon_j$ from stage $t$ to $t+1$ are:
\begin{align*}
\begin{split}
&\epsilon_i^{t+1}=\epsilon_i^{t} + \eta (x_{ij}^{t+1} - \widehat{p}_{ij}^{t}),\\
&\epsilon_j^{t+1}=\epsilon_j^{t} + \eta (x_{ji}^{t+1} - \widehat{p}_{ji}^{t}),
\end{split}
\end{align*}
where $\eta$ is a learning rate ($\eta=16$ or $\eta=32$ in \cite{elo2}), $\widehat{p}_{ij}^{t}:=\sigma(\epsilon_i^{t}-\epsilon_j^{t})$ is the Elo estimate of $\widetilde{P}_{ij}$ at stage $t$, $x_{ij}^{t+1}$ is the outcome of the game between $i$ and $j$ ($x_{ij}^{t+1}$ is 1 if $i$ wins, 0 if $j$ wins, and $\frac{1}{2}$ if $i$ and $j$ draw), so that $x_{ji}^{t+1}=1-x_{ij}^{t+1}$ and $\widehat{p}_{ji}^{t}=1-\widehat{p}_{ij}^{t}$. Note that $\widetilde{P}_{ij} =  \mathbb{E}[x_{ij}^{t+1}]$. The Hyperbolic Elo rating computes the Elo rating of $\varphi_{\beta}(P)$, remembering that $P=2\widetilde{P}-1$. Therefore, the Elo online update rule should be modified as follows:
\begin{align*}
\begin{split}
&\epsilon_i^{t+1}=\epsilon_i^{t} + \eta (f^{t+1}_{ij}(x_{ij}^{t+1}) - \widehat{p}_{ij}^{t}),\\
&\epsilon_j^{t+1}=\epsilon_j^{t} + \eta (f^{t+1}_{ji}(x_{ji}^{t+1}) - \widehat{p}_{ji}^{t}),
\end{split}
\end{align*}
where $f^{t+1}_{ij}(x_{ij}^{t+1})$ is a random variable such that $\EE[f^{t+1}_{ij}(x_{ij}^{t+1})]=\frac{1}{2}(1+\varphi_{\beta}(2\widetilde{P}_{ij}-1))$. Note that as $\beta \to 0$, $\varphi_{\beta} \to \id$ and therefore $\EE[f^{t+1}_{ij}(x_{ij}^{t+1})] \to \widetilde{P}_{ij}$, so that one can take $f^{t+1}_{ij}=\id$, i.e. one recovers the Elo update rule. When $\beta > 0$, however, one needs to keep track of the empirical average of game outcomes $\widetilde{P}_{ij}^t := \frac{1}{|T(i,j,t)|} \sum_{s \in T(i,j,t)} x_{ij}^{s}$, where $T(i,j,t)$ is the set of times $s \in [1,t]$ where $i$ played against $j$. Then, let:
\begin{align*}
\begin{split}
& f^{t+1}_{ij}(x) := \frac{1}{2} + g(x) + \delta_{ij}^{t+1},\\
& \delta_{ij}^{t+1}:= \frac{1}{2}\varphi_{\beta}(2\widetilde{P}_{ij}^{t+1}-1) - g(\widetilde{P}_{ij}^{t+1}),\\
&g(x) := \varphi_\beta(1) (x-\frac{1}{2}).
\end{split}
\end{align*}
By the strong law of large numbers, $\widetilde{P}_{ij}^t$ converges to the constant $\widetilde{P}_{ij}$ almost surely as $t \to +\infty$, and therefore $\EE[f^{t+1}_{ij}(x_{ij}^{t+1})]$ will converge to the desired $\frac{1}{2}(1+\varphi_{\beta}(2\widetilde{P}_{ij}-1))$. Rigorously, one should use a two-timescale framework where $\widetilde{P}_{ij}^t$ is updated on the fast timescale, and $\epsilon_i^{t}$ on the slow timescale so that the former can be approximated by the constant $\widetilde{P}_{ij}$ in the update rule of $\epsilon_i^{t}$, cf. \cite{borkar}.

We made this specific choice for $f^{t+1}_{ij}$ because we want the correction term $\delta_{ij}^{t+1}$ to be as small as possible, since the empirical average $\widetilde{P}_{ij}^{t+1}$ only appears in $\delta_{ij}^{t+1}$, not in $g$. If we could, we would choose $g$ such that $\delta_{ij}^{t+1}=0$ while preserving the constraint on $\EE[f^{t+1}_{ij}(x_{ij}^{t+1})]$, but it is not possible to the best of our knowledge. Another possible choice would have been $g(x)=x$, which would also satisfy that $\EE[f^{t+1}_{ij}(x_{ij}^{t+1})]$ converges to the desired $\frac{1}{2}(1+\varphi_{\beta}(2\widetilde{P}_{ij}-1))$ as $t \to +\infty$. We have checked empirically on a few toy examples that our choice of $g$ yielded smoother and more stable trajectories for the ratings than $g(x)=x$, see Figure~\ref{elo_online} in the Appendix. The heuristic explanation is as follows: a reasonable choice for $f^{t+1}_{ij}$ would have been $f^{t+1}_{ij}(x)=\frac{1}{2}(1+\varphi_{\beta}(2x-1))$. However $\EE[\varphi_{\beta}(2x_{ij}^{t+1}-1)] \neq \varphi_{\beta}(2\EE[x_{ij}^{t+1}]-1) = \varphi_{\beta}(2\widetilde{P}_{ij}-1)$, so we need to correct for the corresponding difference. We get that $\EE[\frac{1}{2}\varphi_{\beta}(2x_{ij}^{t+1}-1)]=\varphi_\beta(1) (\widetilde{P}_{ij}-\frac{1}{2})=g(\widetilde{P}_{ij})$ if the probability that $i$ and $j$ draw is zero. This justifies the choice of this specific form for $g$, where $x$ is scaled by $\varphi_\beta(1)$.

We leave for future work the detailed analysis of such an online update rule, as well as the question of the optimality of our choice of $f^{t+1}_{ij}$. Note that the update rule that we presented is valid not only in the Hyperbolic Elo case where $\varphi_\beta(x)=\frac{1}{\beta} \tanh(\beta x)$, but also for any basis function $\varphi$, in particular if it has been learnt using the methods presented in this work.

\section*{Acknowledgements}
Rahul Savani was supported by a J.P.Morgan AI Research Faculty Award,  the
``Automated Analysis of Strategic Interactions'' project at the Alan Turing Institute, and EPSRC
grant EP/W014750/1.

\section*{Disclaimer}
This paper was prepared for information purposes by the Artificial Intelligence Research group of JPMorgan Chase \& Co and its affiliates (“JP Morgan”), and is not a product of the Research Department of JP Morgan. JP Morgan makes no representation and warranty whatsoever and disclaims all liability, for the completeness, accuracy or reliability of the information contained herein. This document is not intended as investment research or investment advice, or a recommendation, offer or solicitation for the purchase or sale of any security, financial instrument, financial product or service, or to be used in any way for evaluating the merits of participating in any transaction, and shall not constitute a solicitation under any jurisdiction or to any person, if such solicitation under such jurisdiction or to such person would be unlawful.

\bibliography{neurips_2023}
\bibliographystyle{apalike}

\clearpage

\appendix

\onecolumn
\section{EXPERIMENTAL DETAILS AND ADDITIONAL EXPERIMENTS}
\label{app:A}

\subsection{Loss function} 
\label{app:loss}

Let $\DD=\T+\C$ our disk decomposition. Our loss function is:
\begin{align}
\mathcal{L} = \mathcal{L}_{proba} + \omega^{\T}_{sign} \mathcal{L}^{\T}_{sign} + \omega^{\C}_{sign}\mathcal{L}^{\C}_{sign} + \omega_{basis}\mathcal{L}_{basis}.
\end{align}

We take $\omega^{\T}_{sign}=\omega_{basis}=1000$, $\omega^{\C}_{sign}=10$. Let us denote $J_{train}$ the training set, $J^0_{train}$ the set of points $(i,j)$ in the training set such that $P_{ij}=0$. Let $\varphi_m(\DD)$ the $m^{th}$ output of the basis network. For every $(i,j)$, let $m_{ij}=\argmin_m |\varphi_m(\DD_{ij}) - P_{ij}|$. 
\begin{align}
\label{lproba}
\mathcal{L}_{proba} = \frac{1}{4|J_{train}|} \sum_{(i,j) \in J_{train}} ||P_{ij}-\varphi_{m_{ij}}(\DD_{ij})||^2_2.
\end{align}
Let $\rho$ a permutation, and $P_{ij}^\rho$, $\DD^\rho_{ij}$ the corresponding permuted quantities. We want the functions $\varphi_m$ to be nondecreasing, so we define:
\begin{align}
& \mathcal{L}_{basis} = \frac{1}{N_{basis}} \sum_{(i,j) \in J_{train}}\sum_{m=1}^M 
\max\left[0, -(\DD_{ij}-\DD^\rho_{ij})(\varphi_m(\DD_{ij})-\varphi_m(\DD^\rho_{ij}))\right], \\   
& N_{basis} = \frac{1}{4|J_{train}|} \sum_{(i,j) \in J_{train}} |\DD_{ij}-\DD^\rho_{ij}|   \sum_{(i,j) \in J_{train}} \sum_{m=1}^M |\varphi_m(\DD_{ij})-\varphi_m(\DD^\rho_{ij}))|.
\end{align}
The latter can be viewed as similar to the Pearson correlation coefficient. Similarly we have:
\begin{align}
\label{lsignt}
& \mathcal{L}^{\T}_{sign} = \frac{1}{N_{sign}} \left(\sum_{(i,j) \in J_{train}} 
\max\left[0, -\T_{ij} P_{ij}\right] + \sum_{(i,j) \in J^0_{train}} \T_{ij}^2\right), \\
\label{lsignc}
& \mathcal{L}^{\C}_{sign} = \frac{1}{N_{sign}} \left(\sum_{(i,j) \in J_{train}} 
\max\left[0, -\C_{ij} P_{ij}\right]+ \sum_{(i,j) \in J^0_{train}} \C_{ij}^2 \right),\\
& N_{sign} = \frac{1}{|J_{train}|} \sum_{(i,j) \in J_{train}} |\DD_{ij}| \sum_{(i,j) \in J_{train}} |P_{ij}|.
\end{align}
The first term in the latter equations ensures that the sign of $\T$ and $\C$ is that of $P$, the second term make sure that the ties are captured correctly (i.e. the points where $P_{ij}=0$).

\subsection{Game of Figure 1 and additional examples of transitive games} 
\label{app:fig1}

\begin{figure}[ht]
\newcommand{\fsize}{0.9}
  \centering
  \includegraphics[width=0.49\linewidth]{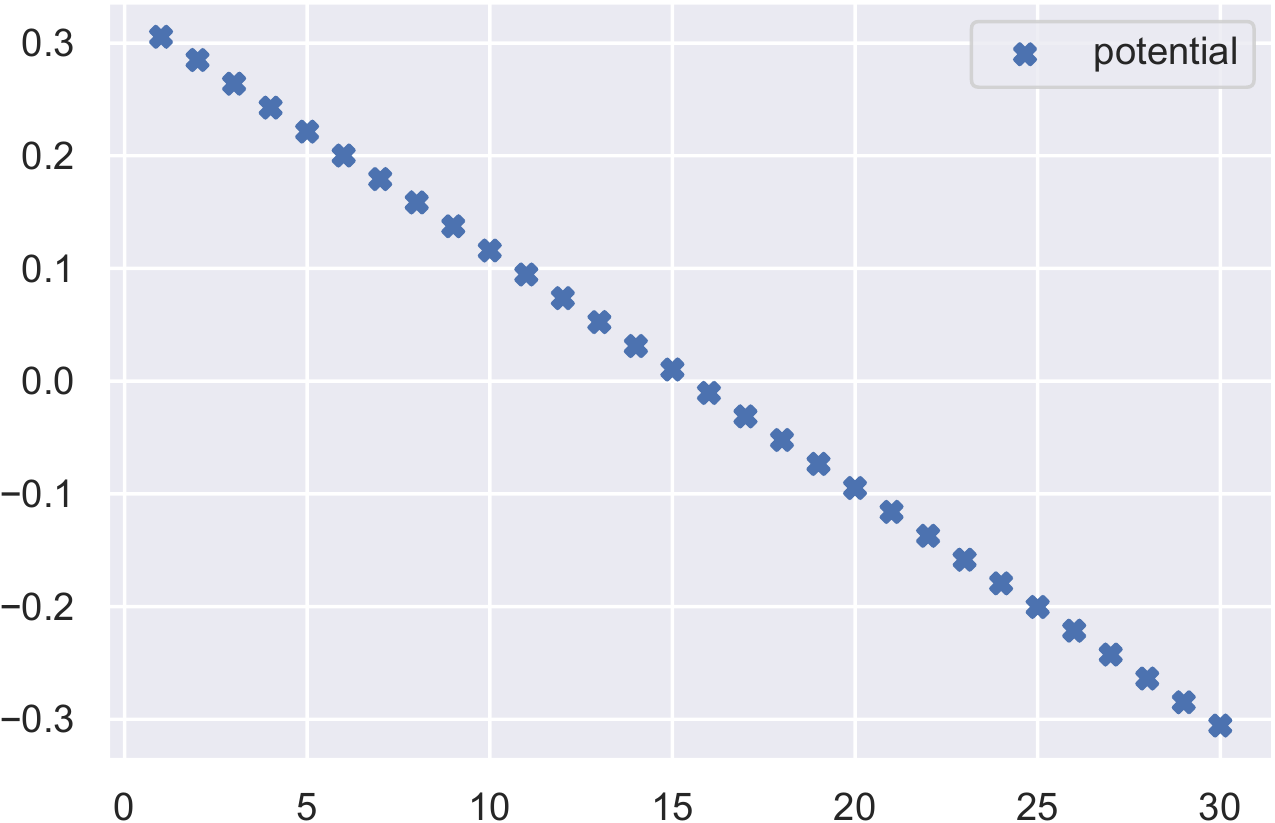}
  \includegraphics[width=0.49\linewidth]{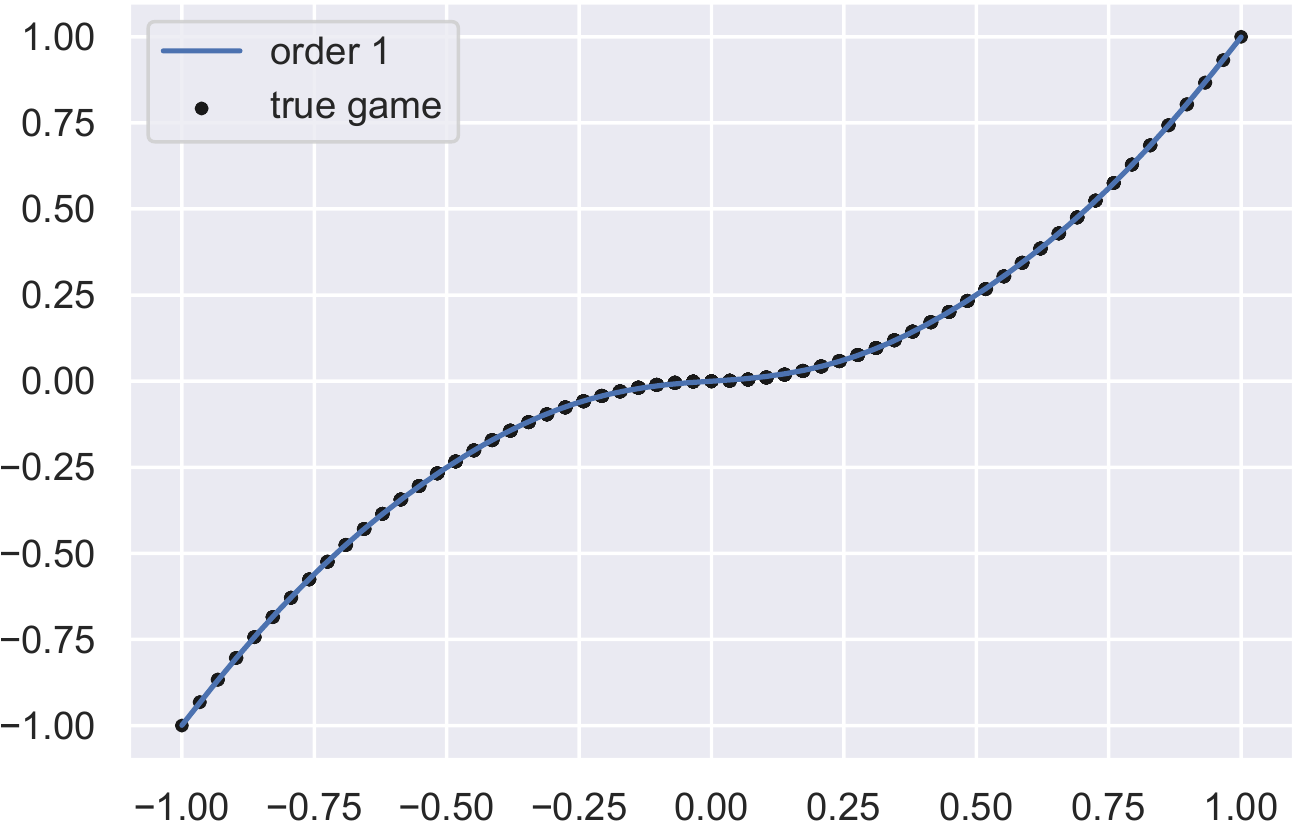}
	\includegraphics[width=0.95\linewidth]{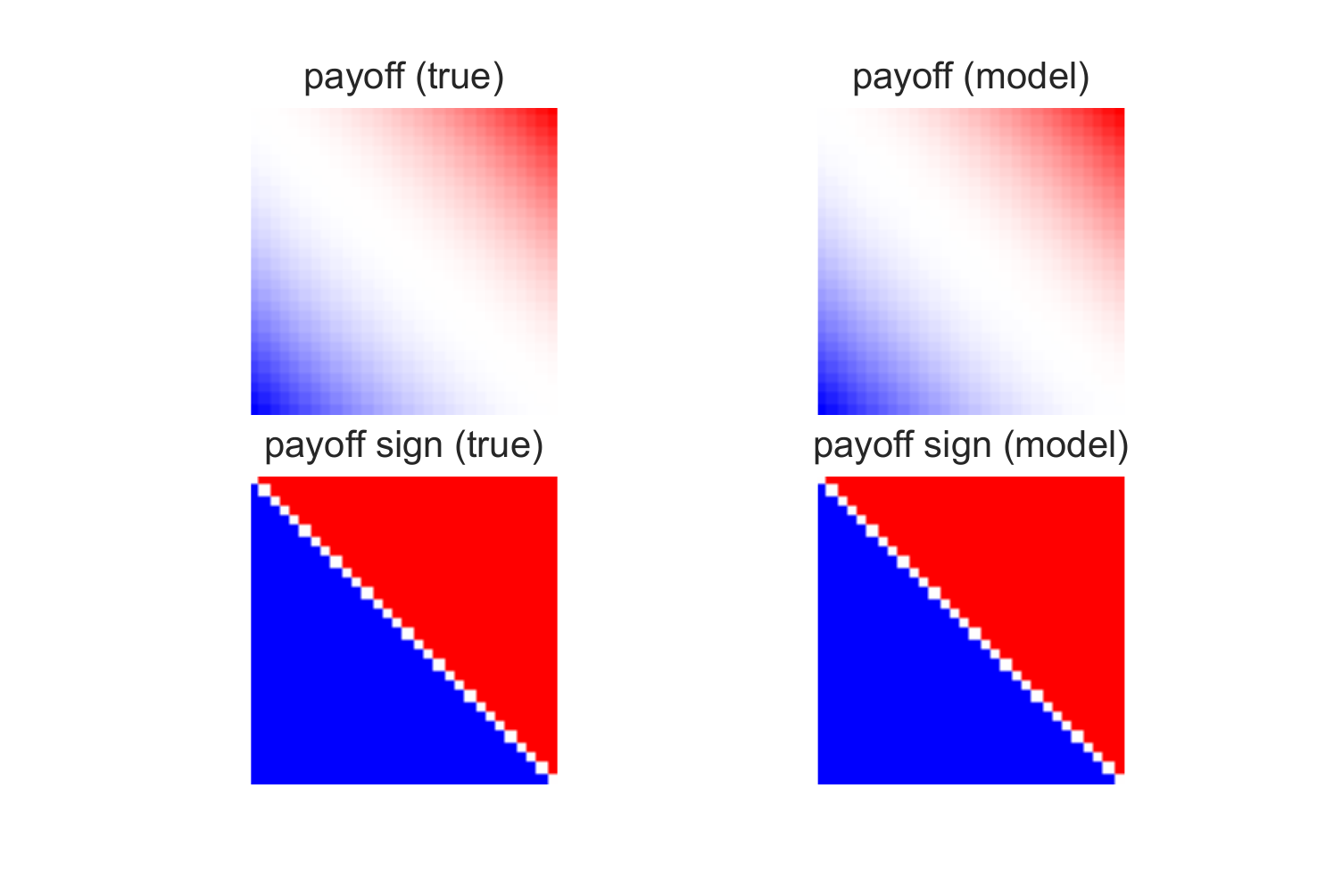}
  \caption{Transitive game of order one of polynomial type, $n=30$. (top left) ordinal potential player scores $\Phi_i$; (top right) game $P$ and its learnt basis function as a function of its disk space representation as in Figure 3; (bottom) learnt and true payoff $P$. We are able to learn that the game is generated by a polynomial function and that the player scores $\Phi_i$ are evenly spaced.}
  \label{fig_t1}
\end{figure}

We consider, for $n=30$, a transitive game of order one of polynomial type, namely $P_{ij} = \varphi(\Phi_i - \Phi_j)$, $\varphi(x):=\lambda \sign(x) \cdot  x^2$, $\Phi_{n-i+1}:=-1 + \frac{2}{n-1} (i-1)$ for $i \in [1,n]$ and $\lambda=0.25$ is a normalization constant. We present in Figure~\ref{fig_t1} the learnt game. We are able to recover the game perfectly, in particular the generating function $\varphi$ and the potential scores $\Phi_i$, evenly spaced. The plot on the top left represents the learnt $\Phi_i$, $i \in [1,n]$. The plot on the top right is similar to Figure 3 and represents the learnt function $\varphi$ (in blue) as well as the points of coordinate $(\DD_{ij}, P_{ij})$. If we learn the game perfectly, the latter points should be on the curve $\varphi$.

\begin{figure}[ht]
\newcommand{\fsize}{0.9}
  \centering
  \includegraphics[width=0.49\linewidth]{figs/trans2/01_potential_order.pdf}
\includegraphics[width=0.49\linewidth]{figs/trans2/02_transitivity_order.pdf}
  \includegraphics[width=0.95\linewidth]{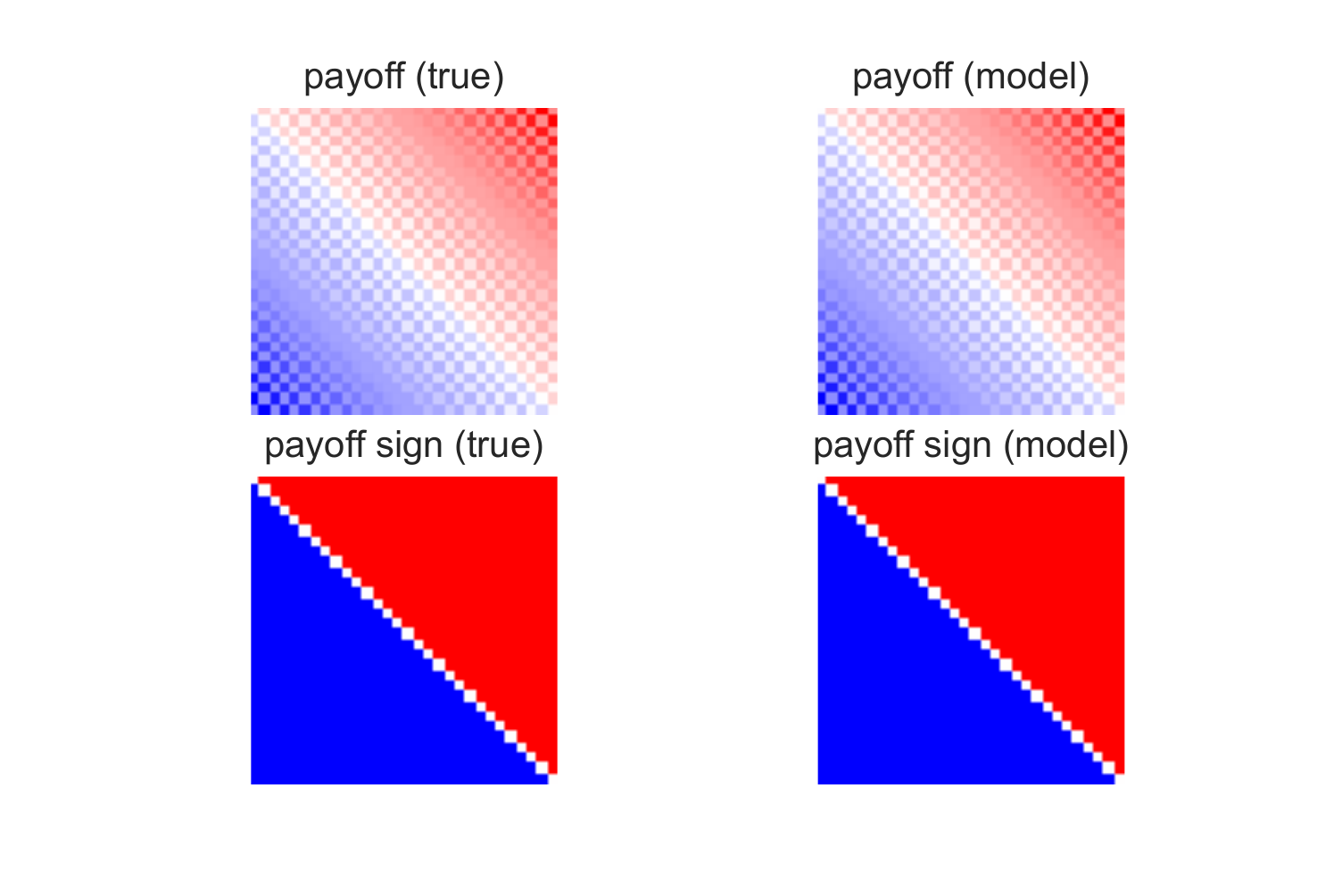}
  \caption{Transitive game of order two of polynomial type, $n=30$. (top left) ordinal potential player scores $\Phi_i$; (top right) game $P$ and its learnt basis functions as a function of its disk space representation as in Figure 3; (bottom) learnt and true payoff $P$. We are able to learn that the game is generated by two polynomial functions and that the player scores $\Phi_i$ are evenly spaced.}
  \label{fig_t22}
\end{figure}

Then, we consider the same setting but now $P_{ij} = \varphi_{ij}(\Phi_i - \Phi_j)$, $\varphi_{ij}(x):=\lambda \sign(x) \cdot  |x|^{1.5}$ if $i+j$ is even, $\varphi_{ij}(x):=\lambda \sign(x) \cdot  |x|^{0.3}$ if $i+j$ is odd and $\lambda=2.7$. Therefore the game is transitive of order two. In Figures~1 and~\ref{fig_t22} we show that we are able to learn the game.

\subsection{Game of Figure 2} 
\label{app:fig2}

\begin{figure}[ht]
\newcommand{\fsize}{0.75}
  \centering
  \includegraphics[width=\fsize\linewidth]{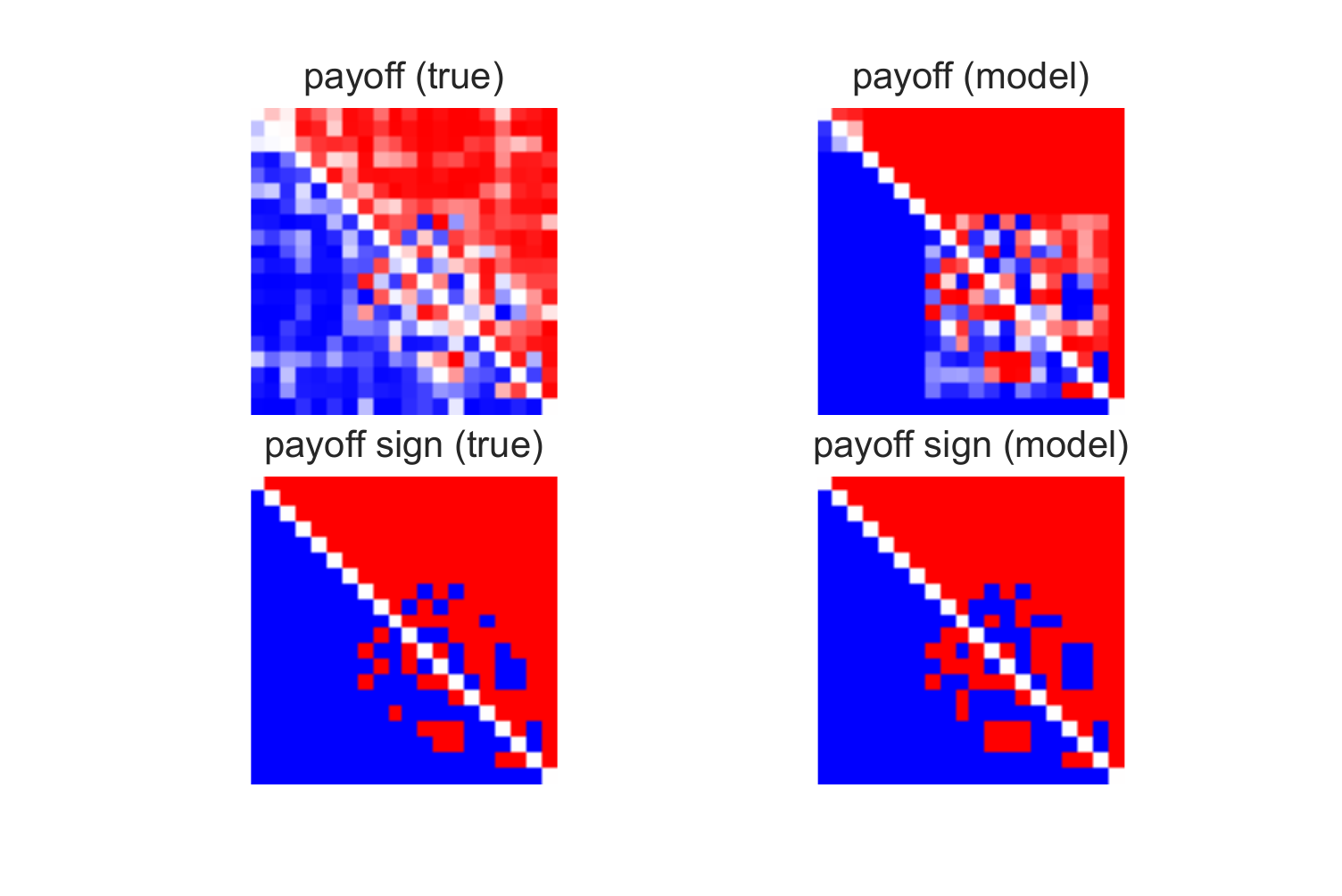}
\includegraphics[width=\fsize\linewidth]{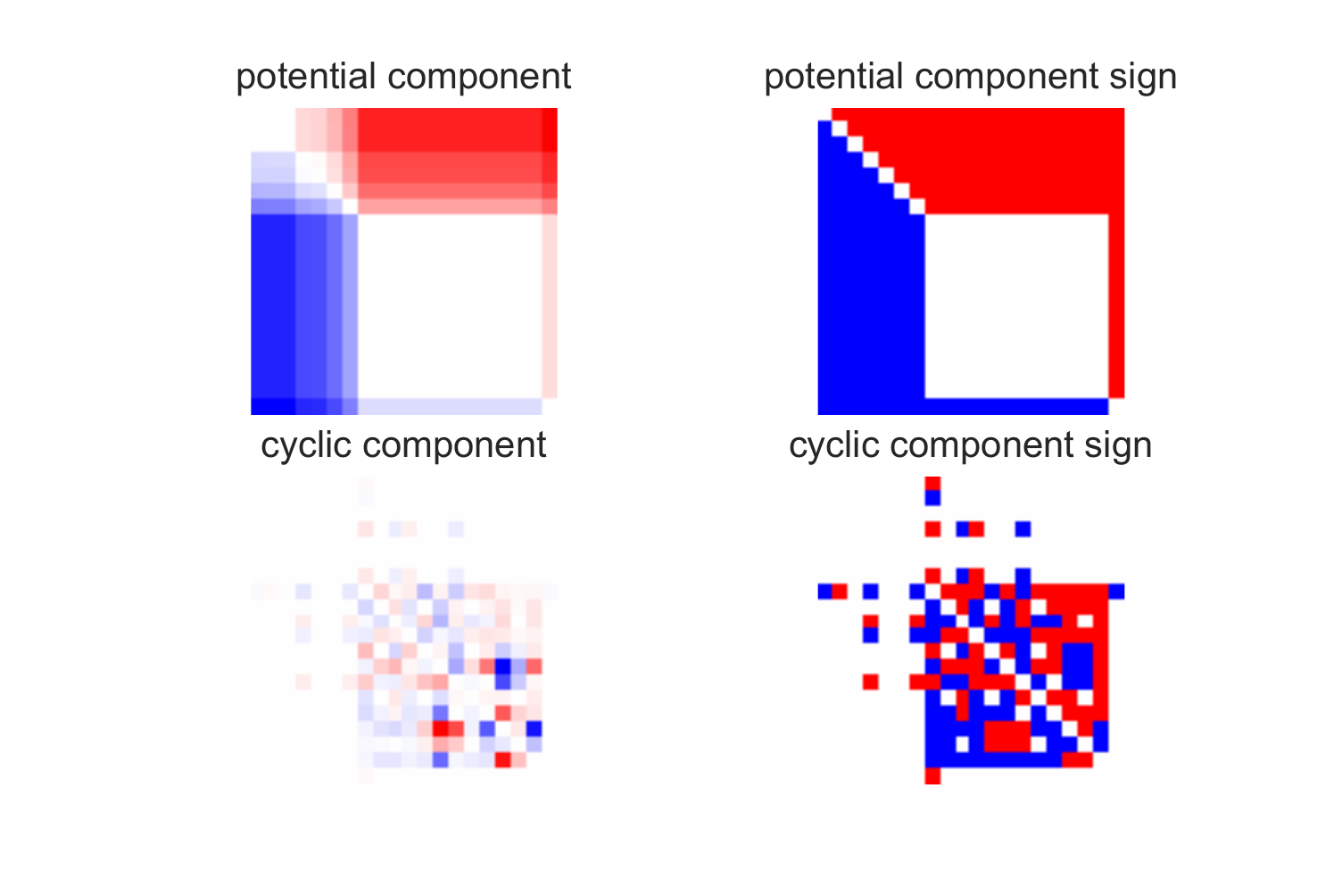}
\includegraphics[width=0.4\linewidth]{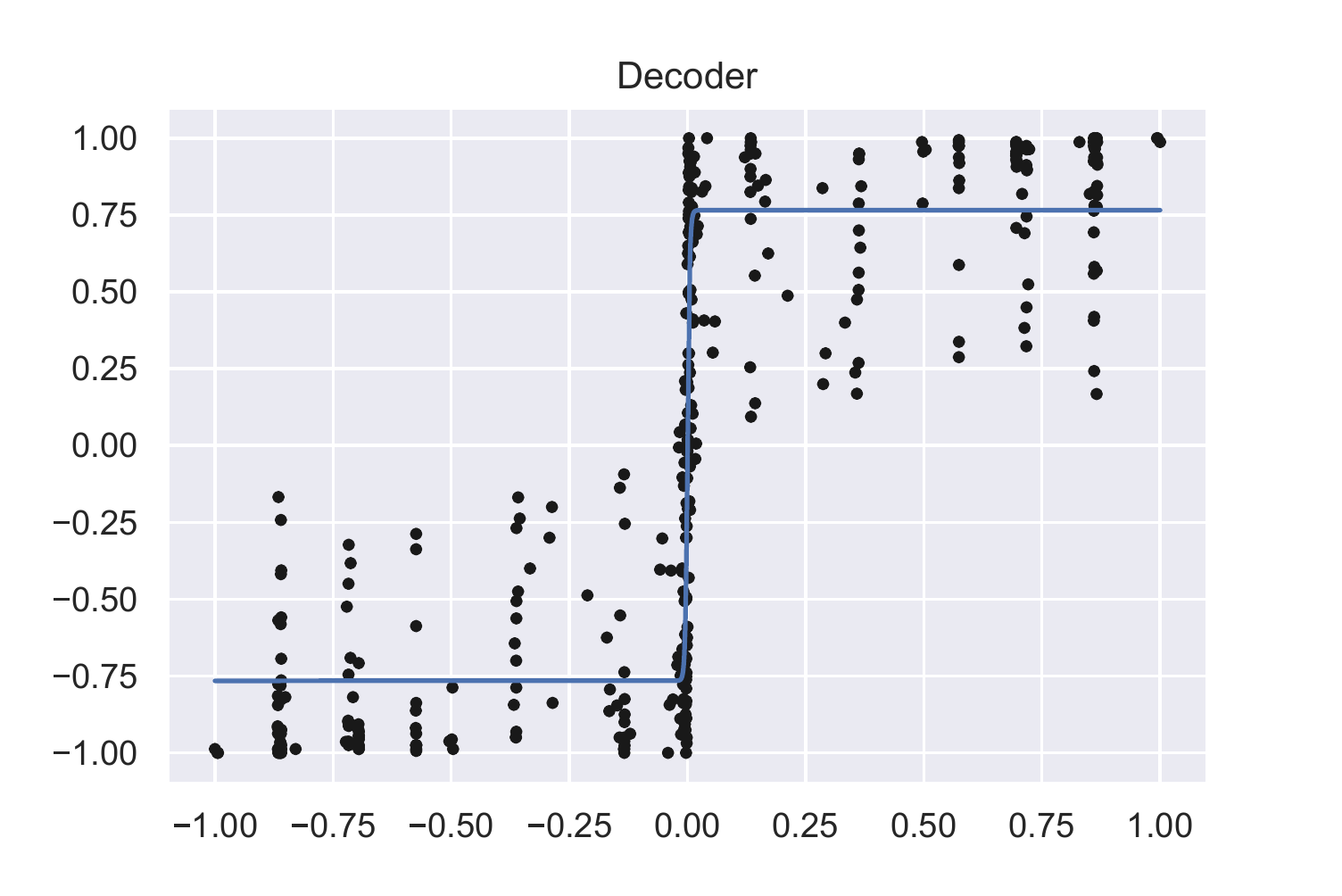}
  \caption{Game of Figure 2: AlphaStar, $n=20$. $K=2$ cyclic components, $M=1$ basis function. Learnt and true Payoff $P$, cyclic component $\C$, transitive (potential) component $\T$.}
  \label{fig_alphas_tc_2}
\end{figure}

To learn the ratings $\Phi_i$, we employ the methodology detailed in Section 3, in particular the architecture in Figure~5. Note that we do note force the 12 players that are part of the large cycle to have the same rating, it is the consequence of our loss function that requires the transitive component $\T$ to have the same sign as $P$, namely $\T_{ij} P_{ij} \geq 0$. We display in Figure~\ref{fig_alphas_tc_2} the true and learnt payoff $P$ (as well as its sign), together with the transitive component $\T$ and cyclic component $\C$. Here, we chose $M=1$ basis function, $K=2$ cyclic component, and a transitive component $\T=Disk(\Phi,\bm{1})$, i.e. $v^\T=\bm{1}$ and $\T_{ij} = \Phi_i - \Phi_j$.

\subsection{Game of Figure 3} 
\label{app:fig3}

We consider the cyclic game given by:
\begin{align*}
& P_{ij} = \lambda \varphi_{ij}(u_i v_j - v_i u_j),\\
&u=(0.16, -0.73,  0.53,  0.22,  0.26,  0.46,  0.35,  0.54, -0.53, -0.05),\\
&v=(-0.39,  0.4 , -0.43, -0.92,  0.31, -0.48, -0.12,  0.38,  0.6 , 0.67),
\end{align*}
where $\lambda=0.72$ is a normalization constant and $\varphi_{ij}(x)=\sign(x) \cdot \sqrt{|x|}$ if $i+j$ is odd,  $\varphi_{ij}(x)=\sign(x) \cdot  x^2$ if $i+j$ is even. The sign of the game $P$ is that of $Disk(u,v)$, hence $P$ is cyclic with sign-rank 2. We present in Figure~\ref{fig_cyclic_1_bis} the equivalent of Figure 3, but for the Elo method.

\subsection{Stability of ratings} 

\begin{figure}[ht]
\newcommand{\fsizee}{0.49}
  \centering
  \includegraphics[width=\fsizee\linewidth]{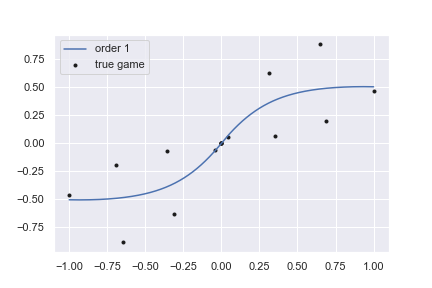}
  \includegraphics[width=\fsizee\linewidth]{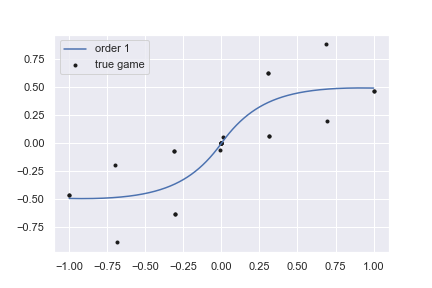}
  \includegraphics[width=\fsizee\linewidth]{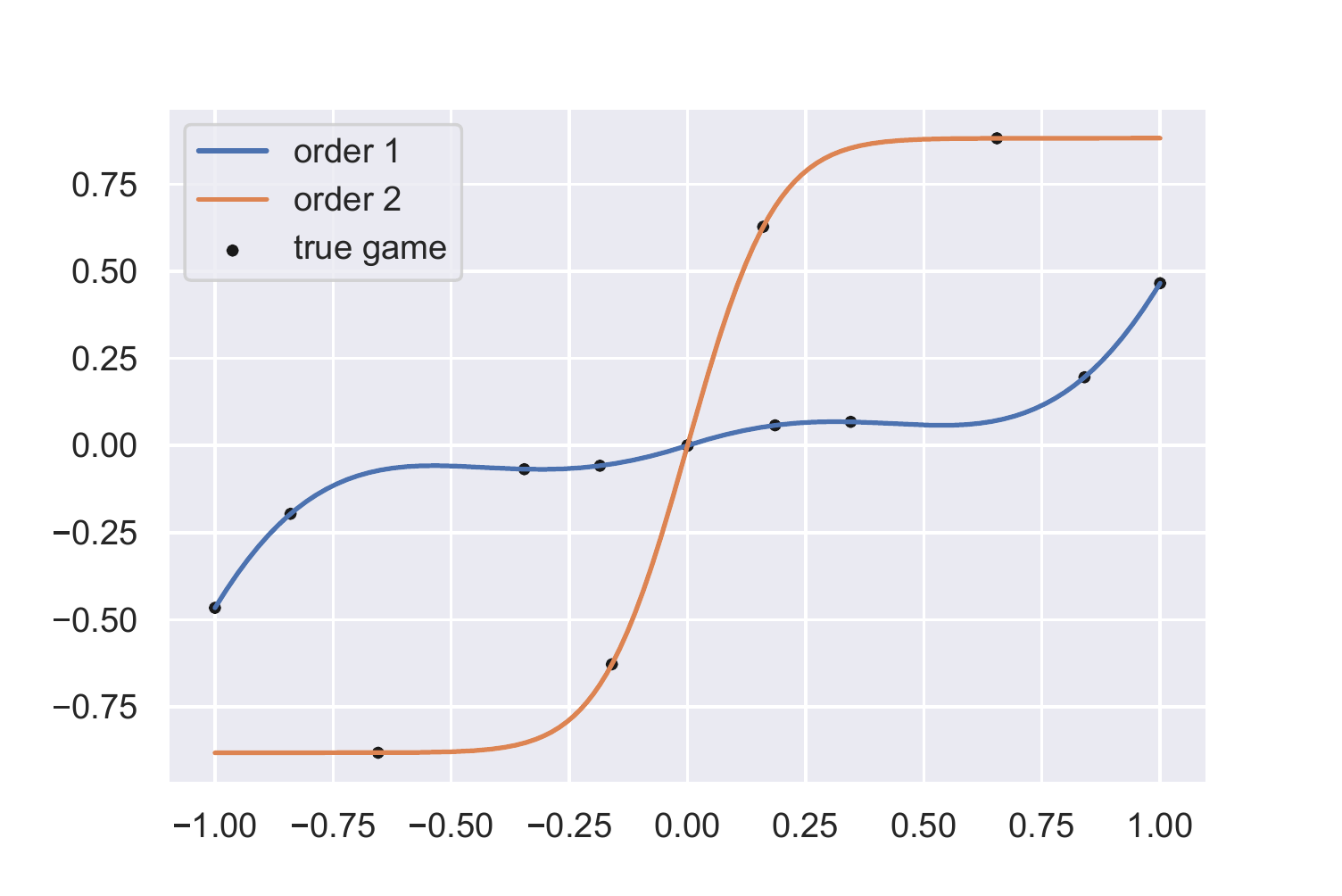}
  \caption{Transitive game of Proposition 1. Our potential-based rating is stable under the addition of a redundant player. (Top left) Four players; (top right) Five players; (bottom) four players, two basis functions (the game is transitive of order two). Black dots represent the true game $P$, the blue and orange curves are the learnt basis function.}
  \label{fig_tran_ex1}
\end{figure}

\begin{figure}[ht]
  \centering
  \includegraphics[width=0.6\linewidth]{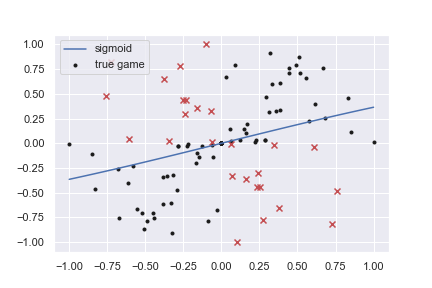}
  \caption{Elo method $\elo(P)$ for the game in Figure 3. As expected, it cannot learn accurately the sign of a cyclic game.}
  \label{fig_cyclic_1_bis}
\end{figure}

It was observed in \cite{BalduzziTPG18} that it is desirable for the rating mechanism to be invariant with respect to the addition of redundant players. Consider the game in (\ref{matpex}), to which we add a copy of player 4. With one basis function, we learn potential ratings of $(1, 0.35, 0.31, 0)$ (4 players) and $(1, 0.31, 0.30, 0, 0)$ (5 players), so the ratings are relatively stable. We show in Figure~\ref{fig_tran_ex1} the learnt basis functions in the two cases. In contrast, the normal decomposition in \cite{BWG23} gives player strength of $(1, 0.22, 0.48, 0)$ and $(1, 0.2, 0.59, 0, 0)$; the Elo ratings are $(1, 0.15, 0.56, 0)$ and $(1, 0.15, 0.66, 0, 0)$. Both these methods see the rating of player 3 vary quite significantly. Note that in all cases, we apply a linear transformation to the ratings so that they lie in $[0,1]$. We believe that the stability of our ratings come from our ability to adjust the amplitude of the reconstructed $\widehat{P}$ with the learnt basis function, whereas in the case of Elo and of the normal decomposition, the basis function is fixed to the sigmoid. Mathematically, we can completely eliminate the impact of redundant players by considering only the unique pairs $(\DD_{ij},P_{ij})$ in the reconstruction loss (\ref{lproba}), as well as the unique pairs $(\T_{ij},P_{ij})$, $(\C_{ij},P_{ij})$ in (\ref{lsignt})-(\ref{lsignc}).

\subsection{Architecture, compute and game data} 

We implement our code in PyTorch. Both our basis and disk network have 3 hidden layers and 200 neurons per layer. All activation functions are tanh, except for the output of the disk network for which the activation function is the identity. We run our experiments on an AWS g3.8xlarge instance, for 60,000 training iterations, with an Adam optimizer with learning rate $5 \cdot 10^{-6}$ ($10^{-4}$ for the first 2,000 iterations). The network weights are initialised using the Xavier (uniform) method. In the computational graph of Figure 5, we perform Gram-Schmidt orthogonalization to the output of the Disk network: we require $v^\T$ to be orthogonal to $u^\T$ and to all the $u^k$'s and $v^k$'s. The $u^k$'s and $v^k$'s are also made orthogonal to each other. To make two vectors $u$ and $v$ orthogonal, we perform:
$$
v \leftarrow v - \frac{\left<u,v\right>}{\max(\left<u,u\right>,\delta)}u, \hspace{4mm} \delta = 10^{-14}.
$$

The game data (i.e., matrices $P$) is taken from the supplementary material of \cite{CzarneckiGTTOBJ20} \footnote{https://proceedings.neurips.cc/paper/2020/hash/ca172e964907a97d5ebd876bfdd4adbd-Abstract.html}.

\subsection{Baselines} 

The baselines Elo, $m$-Elo and the normal decomposition are taken from \cite{BWG23} \footnote{https://github.com/QB3/discrating} (see their appendix A.2). Precisely, let $\C:=\sum_{k=1}^{K} Disk(u^k,v^k)$. The normal decomposition $\C$ is computed by minimizing $\sum_{(i,j) \in J_{train}} \bce(\widetilde{P}_{ij},\sigma(\C_{ij}))$ under the constraint that all the vectors $u^k$ and $v^k$ are orthogonal to each other, and where we remind that $\bce(y,\hat{y}) := - y \ln \hat{y} -(1-y) \ln(1-\hat{y})$. It has $2K$ parameters per player. Regarding $m$-Elo, let $\bar{P}:= P - Disk(u^\T,\bm{1})$, where $u^\T=\frac{1}{n}P\bm{1}$ is the vector containing the column averages and $Disk(u^\T,\bm{1})$ is the transitive component. Then, the cyclic component $\C$ of the $m$-Elo decomposition is computed by minimizing $\sum_{(i,j) \in J_{train}} ||\bar{P}_{ij} - \C_{ij}||^2$, under the constraint that all the vectors $u^k$ and $v^k$ are orthogonal to each other. The $m$-Elo decomposition is then given by $Disk(u^\T,\bm{1}) + \C$. It has $2K+1$ parameters per player.

\subsection{Experiments in Table 1} 

For each random seed and each game, a subset of $n$ players is randomly selected from the full game matrix ($n=50$, $75$, $100$), so that $P$ is of size $n \times n$. Out of all the games presented, only $5,3$-Blotto (21) and Kuhn-poker (64) have a full game matrix $P$ of size less then $100$, so for these two games the number of players selected is the minimum between the full game size and $n$. The training set $J_{train}$ is created by removing $10\%$ of the off-diagonal elements of the matrix $P$, as in \cite{BWG23}. We average our experiments over 3 random seeds. 

All the methods presented have $K=3$ components. As previously discussed, the normal decomposition has $2K$ parameters per player $i \in [1,n]$, a total of $2Kn$ parameters. $m$-Elo additionally adds a transitive component obtained by averaging the columns of $P$, which yields $2K+1$ parameters per player, although it could be argued that the transitive component is not really "learnt" but can simply be seen as a suitable "renormalization", so we believe it is a fair comparison to the normal decomposition with $2K$ parameters. Similarly, we learn a transitive (potential) component $\T:=Disk(u^\T,v^\T)$. However, for the games considered in Table 1, almost all of them are cyclic, which yields $\T=0$. For this reason, we perform a quick check at the beginning of the learning phase and if the game is cyclic, we set $\T=0$ and do not learn the transitive component. Since the latter happens almost all of the time on these examples, we believe it is also a fair comparison to the other methods. Even when the game is not cyclic, it contains a very large cycle, so the impact of $\T$ is minimal on these examples. For completeness we list here the few cases where the game is not cyclic (14 cases out of the 81 combinations "seed $\times$ $n$ $\times$ game"), and therefore where we learn a transitive component: AlphaStar (seed 1: $n=50,100$, seed 2: $n=50,75$, seed 3: $n=50,75,100$); Connect Four (seed 2: $n=50,75,100$, seed 3: $n=50,75$); Go (seed 2, $n=75$, seed 3: $n=50$).

In Table 1, we report the overall sign accuracy on the train and test sets associated to the nonzero entries of $P$, since all methods struggle to exactly predict a zero (and so the sign accuracy for the zero entries of $P$ is always zero). We present in Table~\ref{tab2} the same metrics but we put in brackets the split "(train, test)". In Table~\ref{tab3}, we present the standard deviation associated to Table~\ref{tab2} across the 9 seed $\times$ $n$ combinations. In Tables~\ref{tabmae},~\ref{tabmaes} we do the same for the mean absolute error (MAE). Note that we learn $m=3$ basis functions, which makes our MAE lower than other methods.

\begin{figure}[ht]
\newcommand{\fsize}{0.6}
  \centering
  \includegraphics[width=\columnwidth]{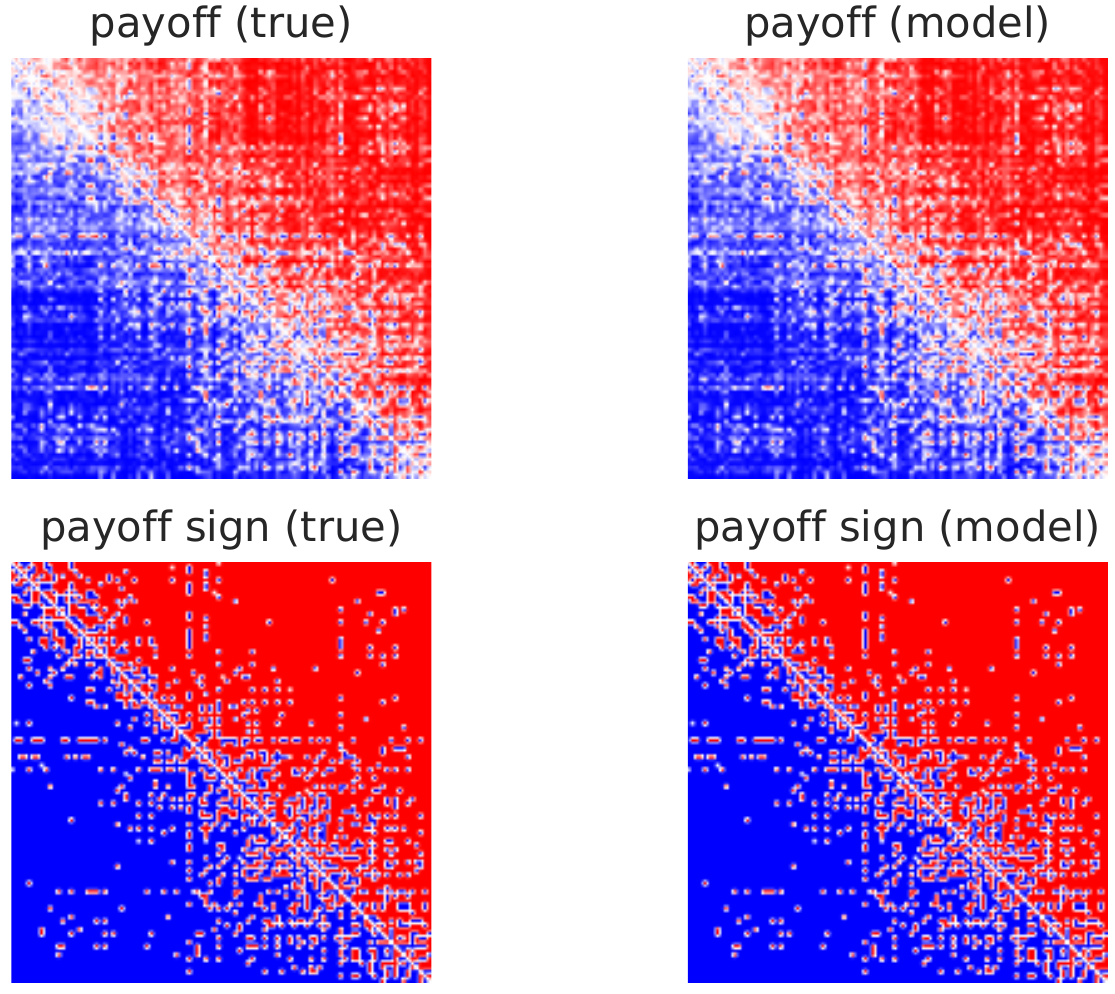}
  \caption{AlphaStar, $n=100$, $M=10$, $K=15$. (top) game payoff; (bottom) game payoff sign; (left) true game; (right) learnt game. Red is positive, blue is negative, white is zero. We train on the full game (no hidden entries), and obtain a sign accuracy of 100\%.}
  \label{figr1}
\end{figure}

Our algorithm is relatively scalable with larger values of $K$, for a fixed game size $n$. Indeed, the only thing that changes when $K$ increases is the size of the output of the disk network $\mathbb{R}^{2K+1}$. In practice, we found that increasing $K$ was not too harmful regarding running time. We provide in Figure~\ref{figr1} an additional experiment in the case of AlphaStar, $n=100$, $K=15$ cyclic components, $M=10$ basis functions.

\section{PROOFS AND TECHNICAL COMMENTS}
\label{app:B}

\begin{remark}
\label{remarkreg} \textbf{(Regular games)}
We have chosen to introduce regular games mainly for clarity of presentation. If ties are allowed, our results would require a slight strengthening of the definition of transitivity, namely that the following additional condition holds: $P_{ij}>0$ and $P_{jk}=0$ implies $P_{ik} \geq0$. This is a natural condition that states that if $A$ wins against $B$, and $B$ ties with $C$, $C$ cannot win against $A$. Then, Theorem 1 would yield that $P_{ij}>0 \Rightarrow \elo(P)_{ij}>0$ $\forall i,j$, namely that Elo preserves transitivity. We commented on this aspect in the proof. Since $P=-P^T$, we also get $P_{ij}<0 \Rightarrow \elo(P)_{ij}<0$. However due to ties, the converse is not true in general, namely when $P_{ij}=0$, we could have either $\elo(P)_{ij}>0$ or $\elo(P)_{ij} \leq 0$. Another way to see it is that $P \sim \elo(P)$ only holds on the set $\mathcal{I}:=\{(i,j): P_{ij} \neq 0\}$. Similarly, the ordinal potential relation in Theorem~\ref{theorempotential} would also hold only on $\mathcal{I}$. 
\end{remark} 

\begin{remark}
\label{remarkbeta} The main merit of the formulas presented in Theorem 1 is that they are explicit. In practice, it is possible to get tighter bounds. Precisely, let:
\begin{align*}
&  P_*:= \max_i \sum_{k=1}^n P_{ik}, \hspace{3mm} P_{**} := - \frac{2n}{3} \arctanh(P_*)^3 + \min_{(i,j): P_{ij}>0} \sum_{k=1}^n P_{ik} - P_{jk}.
\end{align*}
Then $P \sim \elo(P)$ provided $P_*<1$ and $P_{**}>0$. This follows from the proof of Theorem 1 and can be used in conjunction with a one-dimensional root solver to get a tighter lower bound for $\beta$ in Theorem 1. There, one first computes a lower bound $\beta_{low}$ for $\beta$ by solving $P_*=1$, then one finds $\beta>\beta_{low}$ such that $P_{**}=0$.
\end{remark} 

\subsection{Proof of Proposition~\ref{counterex}}
\label{app:proofprop1}
Consider the transitive game:
\begin{align}
\label{matpex}
P = \begin{pmatrix} 
0  &  0.88&  0.2 &  0.46  \\ 
-0.88&  0  & 0.06&  0.06 \\ 
-0.2 & -0.06&  0  &  0.62 \\
-0.46& -0.06& -0.62&  0 \\
\end{pmatrix}.
\end{align}
The Elo ratings are $(0.87, -0.42, 0.19, -0.64)$, and the players' "strength" and "consistency" building the transitive component of the normal decomposition of $\logit \widetilde{P}$ \cite{BWG23} \footnote{They also consider the normal decomposition of $P$ instead of $\logit \widetilde{P}$, but the finding is the same, transitivity is not preserved.} are $(2.66, -1.05, 0.17, -2.04)$ and $(0.67, 0.94, 0.34, 0.38)$, yielding the respective approximations $\elo(P)$ and $\widehat{P}_{\text{NormD}}$:
\begin{align*}
\elo(P) = \begin{pmatrix} 
 0  &  0.57&  0.33&  0.64  \\
-0.57&  0  & \bm{\textcolor{deepcerise}{-0.3}} &  0.11 \\
 -0.33&  \bm{\textcolor{deepcerise}{0.3}} &  0  &  0.39 \\
-0.64& -0.11& -0.39&  0 
\end{pmatrix}, \hspace{5mm} 
\widehat{P}_{\text{NormD}} = \begin{pmatrix} 
 0  &  0.82&  0.27&  0.54 \\
-0.82&  0  & \bm{\textcolor{deepcerise}{-0.19}}&  0.18 \\
-0.27&  \bm{\textcolor{deepcerise}{0.19}}&  0  &  0.14 \\
-0.54& -0.18& -0.14&  0
\end{pmatrix}.
\end{align*}
It is seen that both rating methods do not preserve transitivity of $P$ due to the entries $(2,3)$ of both matrices being negative. In contrast, the Hyperbolic Elo rating in Theorem 1 with $\beta=7$ yields Elo ratings of $\varphi_\beta(P)$ equal to $(0.21, -0.01, -0.02, -0.17)$, and preserves transitivity of $P$:
\begin{align*}
\varphi_\beta^{-1}(\elo(\varphi_\beta(P))) = \begin{pmatrix} 
 0   &  0.148&  0.155&  1  \\
-0.148&  0   &  \bm{\textcolor{darkpastelgreen}{0.003}}&  0.088 \\
 -0.155& \bm{\textcolor{darkpastelgreen}{-0.003}}&  0   &  0.084 \\
-1   & -0.088& -0.084&  0
\end{pmatrix}.
\end{align*}

Similarly, the transitive component of the $m$-Elo method \cite{BalduzziTPG18} is given by $Disk(u^\T,\bm{1})$, where $u^\T$ is the vector containing the column averages $u^\T=\frac{1}{n}P\bm{1}$. It does not preserve transitivity of $P$ and is given by:
\begin{align*}
Disk(\frac{1}{n}P\bm{1},\bm{1}) = 
\begin{pmatrix} 
 0  &  0.57&  0.29&  0.67  \\
-0.57&  0  & \bm{\textcolor{deepcerise}{-0.28}} &  0.1 \\
 -0.29&  \bm{\textcolor{deepcerise}{0.28}} &  0  &  0.38 \\
-0.67& -0.1& -0.38&  0 
\end{pmatrix}.
\end{align*}

Regarding the claim that a transitive game can be decomposed using two cyclic components, let $P$ the regular transitive game:
\begin{align*}
P = \begin{pmatrix} 
 0     &  0.01   &  0.99   &  0.01   &  0.01  \\
-0.01   &  0     &  0.01   &  0.01   &  0.99  \\
 -0.99   & -0.01   &  0     &  0.43&  0.01 \\
-0.01   & -0.01   & -0.43&  0     &  0.99  \\
-0.01   & -0.99   & -0.01   & -0.99   &  0
\end{pmatrix}.
\end{align*}
The normal (real Schur) decomposition of $P$ yields $P=P_1 +P_2$ with:
\begin{align*}
P_1 = \begin{pmatrix} 
 0  &  0.03&  0.15&  0.03& -0.34  \\
-0.03& 0  & -0.35&  0.02&  0.84  \\
-0.15&  0.35& 0  &  0.42&  0.04 \\
-0.03& -0.02& -0.42& 0  &  0.994  \\
0.34& -0.84& -0.04& -0.994& 0
\end{pmatrix},\\
P_2 = \begin{pmatrix} 
 0  & -0.02&  0.84& -0.02&  0.35  \\
0.02& -0  &  0.36& -0.01&  0.15  \\
-0.84& -0.36& 0  &  0.01& -0.03 \\
0.02&  0.01& -0.01& 0  & -0.004   \\
-0.35& -0.15&  0.03&  0.004  & 0
\end{pmatrix}.
\end{align*}
It is easily checked that neither $P_1$ nor $P_2$ is transitive. 

\subsection{Proof of Theorem 1}
\label{app:proofthm1}

We remind that $P = 2\widetilde{P}-1$. By \cite{BalduzziTPG18} (Proposition 1), the Elo ratings $(\varepsilon^P_i)_{i \in [1,n]}$ satisfy:
\begin{align}
\label{eqelo}
\sum_{k=1}^n \sigma(\varepsilon^P_i - \varepsilon^P_k) = \sum_{k=1}^n \widetilde{P}_{ik} \hspace{5mm} \forall i \in [1,n].
\end{align}
The latter is obtained straightforwardly by setting the gradient (with respect to the variables $\varepsilon^P_i$) of the binary cross-entropy loss between $\widetilde{P}_{ij}$ and $\sigma(\varepsilon^P_i - \varepsilon^P_j)$ to zero. Fix a pair $(i,j)$, without loss of generality we assume that $i$ beats $j$, i.e. $P_{ij}>0$. The goal is to show that $\varepsilon^P_i > \varepsilon^P_j$. Denote $f(x):=2\sigma(x)-1=\tanh(\frac{x}{2})$. By (\ref{eqelo}) we get:
\begin{align*}
& \sum_{k=1}^n f(\varepsilon^P_i - \varepsilon^P_k) - f(\varepsilon^P_j - \varepsilon^P_k) = \sum_{k=1}^n P_{ik} - P_{jk}.
\end{align*}
We have the inequality $|f(x)- \frac{x}{2}| \leq \frac{1}{24}|x|^3$ $\forall x$, which yields:
\begin{align*}
\frac{n}{2} (\varepsilon^P_i - \varepsilon^P_j) \geq \sum_{k=1}^n (P_{ik} - P_{jk}) - \frac{n}{12} \varepsilon_{max}^3,
\end{align*}
where $\varepsilon_{max}:= \max_{i,j} |\varepsilon^P_i-\varepsilon^P_j|$. Let us look carefully at the terms $P_{ik} - P_{jk}$ for $k \neq i,j$. By the regularity assumption, $P_{ik}$ and $P_{jk}$ cannot be zero. By the transitivity assumption, remembering that $P_{ij}>0$, if $P_{jk}>0$ then $P_{ik}>0$. So in that case, $P_{ik} - P_{jk} \geq -(P_{max}-P_{min})$. Same conclusion if $P_{jk}<0$ and $P_{ik}<0$, or if $P_{jk}<0$ and $P_{ik}>0$. This means that in all cases, $P_{ik} - P_{jk}$ is lower bounded by $-(P_{max}-P_{min})$. Note that we really need transitivity here to avoid the "bad" case where $P_{jk}>0$ and $P_{ik}<0$, which would yield the "bad" lower bound $-2P_{max}$. 

Since $P_{ij}>0$ by assumption, then $P_{ij} \geq P_{min}$. Therefore, we have overall:
\begin{align*}
\sum_{k=1}^n P_{ik} - P_{jk} \geq n P_{min} - (n-2) P_{max}.
\end{align*}

Note that if we had allowed ties, namely $P_{jk}$ and $P_{ik}$ can be zero, we would need the additional requirement that if $P_{jk}=0$, then $P_{ik} \geq 0$ so that $P_{ik} - P_{jk} \geq 0$. This is the only "bad" case to handle, since the other cases are the same as above. Indeed, if $P_{jk}>0$, we saw that $P_{ik}>0$ by transitivity, and if $P_{jk}<0$, then the worst case is that $P_{ik}<0$ too but even in that case $P_{ik} - P_{jk}$ is lower bounded by $-(P_{max}-P_{min})$.

Let $\delta := \frac{P_{max}}{P_{min}}$. Overall we get:
\begin{align*}
\frac{n}{2 P_{min}} (\varepsilon^P_i - \varepsilon^P_j) \geq n - (n-2) \delta - \frac{n}{12 P_{min}} \varepsilon_{max}^3.
\end{align*}
Let $i_{max} = \argmax_i \varepsilon^P_i$.  Therefore $\varepsilon^P_{i_{max}} - \varepsilon^P_k \geq 0$ $\forall k$, and equation (\ref{eqelo}) gives us:
\begin{align*}
& \sigma(\varepsilon_{max}) - \frac{1}{2} \leq \sum_{k=1}^n \left(\sigma(\varepsilon^P_{i_{max}} - \varepsilon^P_k) - \frac{1}{2} \right)\leq \frac{n-1}{2} P_{max}\\
\Rightarrow & \hspace{3mm} \varepsilon_{max} \leq \sigma^{(-1)} \left(\frac{1}{2} + \frac{n-1}{2} P_{max}\right) = 2 \arctanh \left((n-1) P_{max}\right).
\end{align*}

For $\alpha>0$, let $x_\alpha$ the unique positive root of $2\arctanh^3 \left(x\right)-3\alpha x$, so that $8\arctanh^3 \left(x\right) \leq 12\alpha x$ for $0 \leq x \leq x_\alpha$. For $P_{max}< \frac{x_\alpha}{n-1}$ we have $\varepsilon^P_i-\varepsilon^P_j>0$ provided:
\begin{align*}
 n - \delta \left( n-2 + n (n-1) \alpha\right) >0 \hspace{3mm} \Leftrightarrow \hspace{3mm} \delta < \frac{n}{n-2 + n (n-1) \alpha}.
\end{align*}

Note that $\delta \geq 1$ by construction, which is the reason why we require $0<\alpha<\frac{2}{n(n-1)}$, so that $\frac{n}{n-2 + n (n-1) \alpha}>1$. The requirement $P_{max}< \frac{x_\alpha}{n-1}$ ensures that $\varepsilon_{max}^3 \leq 12 \alpha (n-1) P_{max}$. 

Finally, note that $\varphi_{\beta}(P_{max}) \leq \beta^{-1}$, and that $P_{ij}>0 \Leftrightarrow \varphi_{\beta}(P_{ij})>0$.

\begin{proposition}
\label{wsopprop} A two-player symmetric zero-sum game $P$ is a weak separable ordinal potential game if and only if there exists a vector $\Phi$ such that:
\begin{align}
\label{wsop}
 P_{ij} > 0 \Leftrightarrow \Phi_{i}-\Phi_{j} > 0 \hspace{6mm} \forall i,j.
\end{align}
\end{proposition}

\textbf{Proof}. The result follows immediately from the definition. Indeed, assume $P$ is a weak separable ordinal potential game. Then $\widetilde{\Phi}_{ij}-\widetilde{\Phi}_{jj} = \alpha_{i} -\alpha_{j}$ and $\widetilde{\Phi}_{ji}-\widetilde{\Phi}_{jj} = \beta_{i} -\beta_{j}$. Also note that $P_{jj}=0$ due to $P$ being antisymmetric. Hence $P_{ij} > 0 \Leftrightarrow \Phi_{i}-\Phi_{j} > 0$ with $\Phi:=\alpha$. Conversely, assume that $P_{ij} > 0 \Leftrightarrow \Phi_{i}-\Phi_{j} > 0$. Define $\alpha:=\beta:=\Phi$ and $\widetilde{\Phi}_{ij} := \alpha_{i} + \beta_{j}$, then $\widetilde{\Phi}_{ij}-\widetilde{\Phi}_{jj} = \widetilde{\Phi}_{ji}-\widetilde{\Phi}_{jj} = \Phi_{i}-\Phi_{j} > 0 \Leftrightarrow P_{ij} > 0$.

\subsection{Proof of Theorem~\ref{theorempotential}}
\label{app:proofthm2}

Direction "$\Leftarrow$". By Proposition 2, assume that $\forall i,j$:
\begin{align*}
 P_{ij} > 0 \Leftrightarrow \Phi_{i}-\Phi_{j} > 0.
\end{align*}
Assume $P_{ij} > 0$ and $P_{jk} > 0$. We want to show $P_{ik} > 0$. $P_{ij} > 0$ and $P_{jk} > 0$ so $\Phi_{i}-\Phi_{j} > 0$ and $\Phi_{j}-\Phi_{k}> 0$. Hence $\Phi_{i}-\Phi_{k} = \Phi_{i}-\Phi_{j} + \Phi_{j}-\Phi_{k} > 0$, hence $P_{ik} > 0$ by assumption. Note that we have used both "if" and "only if" directions in the ordinal potential assumption, first to move to the potential, then to move back to the payoff.

Direction "$\Rightarrow$". This direction is the most challenging and is a consequence of Theorem 1. Indeed, by the latter, take $\varphi_{\beta}(x):=\frac{1}{\beta} \tanh(\beta x)$ for large enough $\beta$. Then, $P \sim \elo(\varphi_{\beta}(P))$, which means that the Elo rating $\varepsilon^{\varphi_{\beta}(P)}$ of the game $\varphi_{\beta}(P)$ satisfies $P_{ij}>0\Leftrightarrow \varepsilon^{\varphi_{\beta}(P)}_i - \varepsilon^{\varphi_{\beta}(P)}_j>0$ $\forall i,j$ and is therefore a weak ordinal separable potential by Proposition 2.

\subsection{Proof of Theorem 3} 
\label{app:proofthm3}

Without loss of generality (upon rearranging the player labels $i$) we assume that we have the maximal cycle $\OO := 1 \to 2 \to ... \to n \to 1$, which corresponds to the permutation $\gamma= \id$ in the definition of cyclicity. That is, $P_{i i+1}>0$ $\forall i \in [1,n-1]$ and $P_{n 1}>0$. 

The direction "$\Leftarrow$" is straightforward: if $K$ disks all admit $\OO$ as a cycle, then the entries $(i,i+1)$ and $(n,1)$ have the same positive sign in all disks, in particular $\sum_{k=1}^{K} Disk(u^k,v^k)$ has positive entries $(n,1)$ and $(i,i+1)$ $\forall i \in [1,n-1]$. But by assumption $P$ is equal to $\sum_{k=1}^{K} Disk(u^k,v^k)$, so $P_{i i+1}>0$ and $P_{n 1}>0$, i.e. $P$ admits $\OO$ as a cycle. Since $\OO$ is of length $n$, it is a maximal cycle.

Direction "$\Rightarrow$". For any two vectors $u$, $v$, we can consider the polar coordinate parametrization $u_i = \rho_i \cos(\theta_i)$, $v_i = \rho_i \sin(\theta_i)$, which yields $u_i v_j - u_j v_i = \rho_i \rho_j \sin(\theta_i - \theta_j)$ \cite{BWG23}. Note that $\rho_i = \sqrt{u_i^2 + v_i^2}\geq 0$, $\theta_i \in [0, 2 \pi)$ and $u_i v_j - u_j v_i$ is the $(i,j)$ entry of $Disk(u,v)$.

First, we show that with $2$ disks, we can a) capture the sign of all pairs $(i,n)$ for $i \in [1,n]$, b) each one of the 2 disks captures correctly the sign of adjacent pairs $(i,i+1)$ $i \in [1,n-1]$ and $(1,n)$. 

b) is easy to achieve and ensures that each one of the two disks is cyclic due to $1 \to 2 \to ... \to n \to 1$ by assumption. Indeed, the sign of the pair $(i,j)$ is determined by that of $\sin(\theta_i - \theta_j)$, so to ensure that adjacent pairs have correct sign, we simply need to put the points $i$ one after another on the trigonometric circle, starting from $\theta_1=0$ and going clockwise, with a spacing between each pair $\theta_{i} - \theta_{i+1}$ no more than $\pi$, and with the last point $n$ in the upper-half of the trigonometric circle to ensure $\theta_n - \theta_{1} > 0$. This can always be achieved, for example by taking $\theta_n = \frac{\pi}{2}$, and $\theta_i = -\frac{(i-1)\pi}{n}$ for $i \in [1,n-1]$.

For a), it is a bit more subtle. By the regularity assumption on the game, either $P_{in}>0$ or $P_{in}<0$. We split the players $i$ in 2 groups: those who lose against $n$ ($P_{i n}<0$), and those who win against $n$ ($P_{i n}>0$). Consider 2 disks with player parameters $\rho^{(k)}_i$, $\theta^{(k)}_i$ for $k =1,2$ and $i \in [1,n]$. The idea will be the following: we aim to capture the sign of $P_{i n}$ for players $i$ in the first group with the first disk, so $\rho^{(1)}_i \gg \rho^{(2)}_i$, and we aim to capture the sign of $P_{i n}$ for players $i$ in the second group with the second disk, so $\rho^{(2)}_i \gg \rho^{(1)}_i$. For the first group, we have $P_{i n}<0$, so we take $\theta^{(1)}_1 =0$, $\theta^{(1)}_n =\frac{\pi}{2}$, and $\theta^{(1)}_i \in (-\frac{\pi}{2}, 0)$ for $i \in [2,n-1]$, for example $\theta^{(1)}_i = -\frac{i\pi}{2n}$. Note that this parametrization captures both b), together with the sign of $P_{i n}$ for players in the first group. Similarly, for the second group, we have $P_{i n}>0$, so we take $\theta^{(2)}_1 =0$, $\theta^{(2)}_n =\frac{\pi}{2}$, and $\theta^{(2)}_i \in (-\pi, -\frac{\pi}{2})$ for $i \in [2,n-1]$, for example $\theta^{(2)}_i =-\frac{\pi}{2} -\frac{i\pi}{2n}$. Overall, we take $\rho^{(1)}_n=\rho^{(2)}_n =: \rho_n$ and we capture the sign of $P_{i n}$ using:
\begin{align}
\label{eqsin0}
   [Disk(u^{(1)},v^{(1)})+Disk(u^{(2)},v^{(2)})]_{in} = \rho_n \left(\rho^{(1)}_i \sin(\theta^{(1)}_i - \theta^{(1)}_n) + \rho^{(2)}_i \sin(\theta^{(2)}_i - \theta^{(2)}_n) \right).
\end{align}

For players $i$ in the first group, take:
\begin{align}
\label{eqsin}
    \rho^{(1)}_i > \rho^{(2)}_i \frac{|\sin(\theta^{(2)}_i - \theta^{(2)}_n)|}{|\sin(\theta^{(1)}_i - \theta^{(1)}_n)|} > \rho^{(2)}_i \sin(\frac{\pi}{2n}).
\end{align}
Similarly, for players $i$ in the second group, take $\rho^{(2)}_i > \rho^{(1)}_i \sin(\frac{\pi}{2n})$. These choices makes the sign of (\ref{eqsin0}) equal to that of $P_{i n}$ for all $i$.

We have seen that we can explain the sign of all $P_{in}$ using 2 cyclic disks. Note that $\rho_n$ has been left unspecified. Now that the signs of all interactions of player $n$ have been captured, we are free to pick $\rho_n$ as large as we want for these 2 disks, and $\rho^{(k)}_n$ as small as we want in any further disk $k$ we may want to add to the decomposition, to ensure - as we did in (\ref{eqsin}) - that we do not change the sign of $P_{in}$ by adding further disks. However, we cannot take $\rho^{(k)}_n = 0$ since we need all points on every disk to satisfy condition b) and guarantee cyclicity, so we cannot "eliminate" player $n$ from any disk.

Note that we have shown a way to explain the sign of all $P_{i n}$ using 2 cyclic disks. We can actually do better in some cases, in the sense that we only need one cyclic disk. This is the case if there exists $k_*=k_*(n) \in [2,n-1]$ such that all players $i \in [2,k_*)$ are in the same group, and all players $i \in [k_*,n-1]$ are in the same group. If players $i \in [2,k_*)$ are all in group 1 ($P_{in}<0$) and all players $i \in [k_*,n-1]$ are all in group 2 ($P_{in}>0$), take $\theta_1=0$, $\theta_n=\frac{\pi}{2}$, $\theta_i =-\frac{i\pi}{2n}$ for $i \in [2,k_*)$, $\theta_i =-\frac{\pi}{2} -\frac{(i+2-k_*)\pi}{2n}$ for $i \in [k_*,n-1]$. This satisfies both a) and b). Similarly, if players $i \in [2,k_*)$ are all in group 2 and all players $i \in [k_*,n-1]$ are all in group 1, take $\theta_n= \pi - \frac{\pi}{2n}$ so that $\theta_n - \theta_2 = \pi - \frac{\pi}{2n} + \frac{\pi}{n} = \pi + \frac{\pi}{2n} \in (\pi, 2 \pi)$, and  $\theta_n - \theta_{k_*} \in (0, \pi)$.

To conclude the proof, we reiterate the procedure that we employed to explain the sign of $P_{in}$, but now we apply it to explain the sign of $P_{i n-1}$ for $i \in [1,n-2]$ (since $P_{n n-1}$ has already been explained). As mentioned earlier, for all the subsequent disks that we will add, $\rho^{(k)}_n>0$ will be chosen as small as desired so as not to perturb the sign of the $P_{in}$'s, in other words we will use (\ref{eqsin}) only for the points $i$ that we haven't explained yet, namely $i \in [1,n-2]$. Precisely, we define $\gamma_1$ as the permutation that pushes each player by $1$ unit back, i.e. $\gamma_1(i)=i-1$ and $\gamma_1(1)=n$. Then, we can apply our previous analysis with the points $\gamma_1(i)$ in the role of the points $i$, in particular $n-1$ in the role of $n$, splitting into 2 groups those players who win and lose against $n-1$. Our construction implies that b) is satisfied, hence all these disks are cyclic; it also implies that we add, in the worst case, $2$ disks per iteration. We repeat this procedure, at each stage $p$, to explain the interactions of player $n-p$ with players $i \in [1,n-p-1]$, using the permutation $\gamma_p$ that pushes each player by $p$ units back. If $\rho^{(k,p)}_i$, $\theta^{(k,p)}_i$ are the parameters of players $i$ for the two disks $k$ at stage $p$, our construction is based on the observation that we are free to choose $\rho^{(k,p)}_i$ as small as desired for $i>n-p$. Stage $p=0$ correspond to player $n$, stage $p=1$ to player $n-1$, etc. We stop at stage $p=n-3$. This is because a single disk will always suffice (we have $P_{12}>0$ and $P_{23}>0$ by assumption on the cycle so the only interaction to explain is $P_{13}$). This can be viewed as an "onion" method, always capturing the full cycle and going deeper each stage to explain more of the players interactions inside the full cycle. 

Overall, we have constructed, for each stage $p$, two disks that capture correctly the signs of interactions of player $n-p$ with all players $i \in [1,n-p-1]$. These disks all contain the maximal cycle $\OO$. 

Our proof shows that we have $1 + 2(n-3)$ cyclic disks at most when $n \geq 5$. This is a worst case scenario and in practice, we will need fewer disks. If $n \geq 5$, when we get at stage $p=n-4$ to explain the interactions of player $4$ with $1$, $2$ and $3$, in general we will need two disks because the interactions $(1,4)$ and $(2,4)$ can be arbitrary, and so we cannot always, with one disk, capture correctly these 2 interactions correctly together with the cycle $\OO$. Precisely, in a given disk, if $P_{42}>0$, we must also have $P_{41}>0$ if we want to respect $\OO$. If $n=4$ however, the interaction $(1,4)$ cannot be arbitrary as it is constrained by $\OO$: $P_{41}>0$, so in that case we will always need one disk only.

\textbf{Corollary 3.1. (Number of cyclic components capturing a cyclic game)} Under the setting of Theorem 3, if $n \leq 4 $, we have $K=1$. If $n \geq 5$, we have $K \leq 2(n-3)+1$. Assume without loss of generality that a maximal cycle is $\OO := 1 \to 2 \to ... \to n \to 1$. For a given player $i$, let $J_i$ the set of players excluding the adjacent players $i-1$ and $i+1$ (modulo $n$). Let $n_*$ the number of players $i$ such that there exists a $k_*(i)$ such that all players in $J_i$ whose index is less than $k_*(i)$ win against $i$, and all players in $J_i$ whose index is greater than $k_*(i)$ lose against $i$ (or vice versa). Then, $K \leq 2(n-3)-n_*+1$.

\textbf{Proof.} The proof follows from the argument developed in the proof of Theorem 3. There, it is seen that if $n=4$, we have $K=1$. If $n \geq 5$, we have $K \leq 2(n-3)+1$ since we add at most 2 disks per stage. Under the existence of $k_*$, we add only one disk, hence the result.

%\clearpage

% player subset sorted u = np.array([0,3,7,8,9,10,11,420,430,440,450,460,470,480,490,500,510,520,530,800])

\begin{figure}[ht]
\newcommand{\fsize}{0.9}
  \centering
  \includegraphics[scale=0.58]{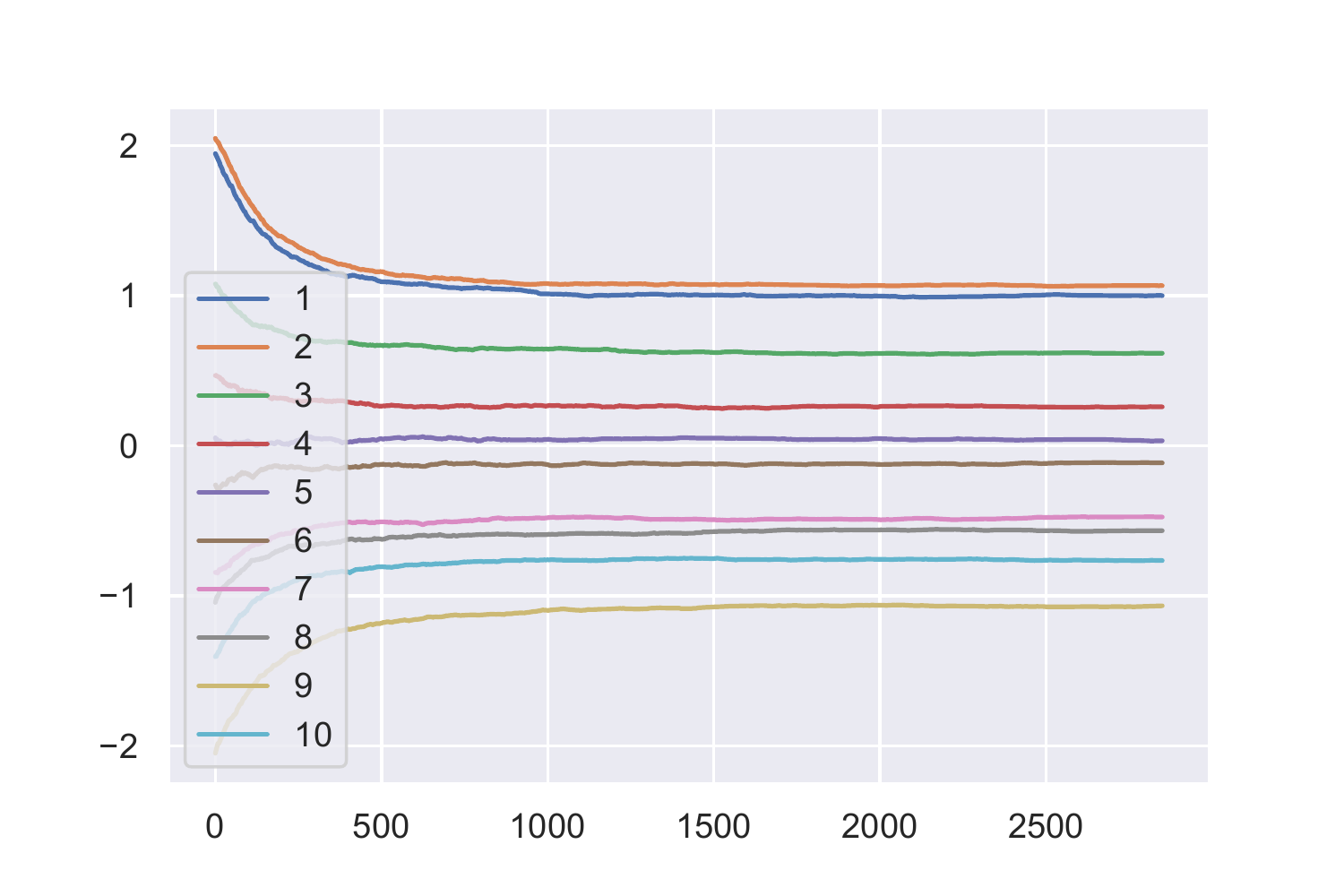}
  \includegraphics[scale=0.58]{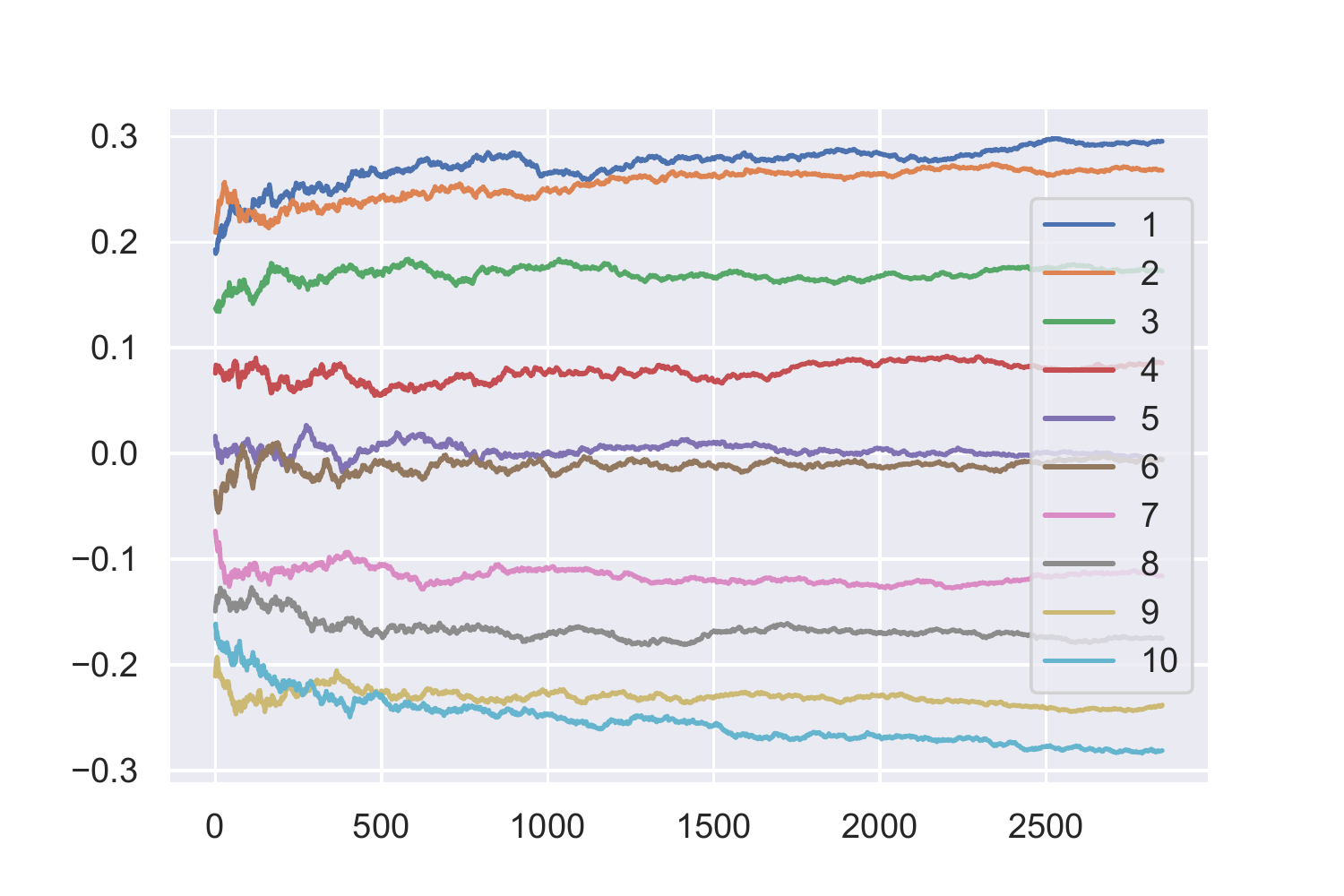}
  \includegraphics[scale=0.58]{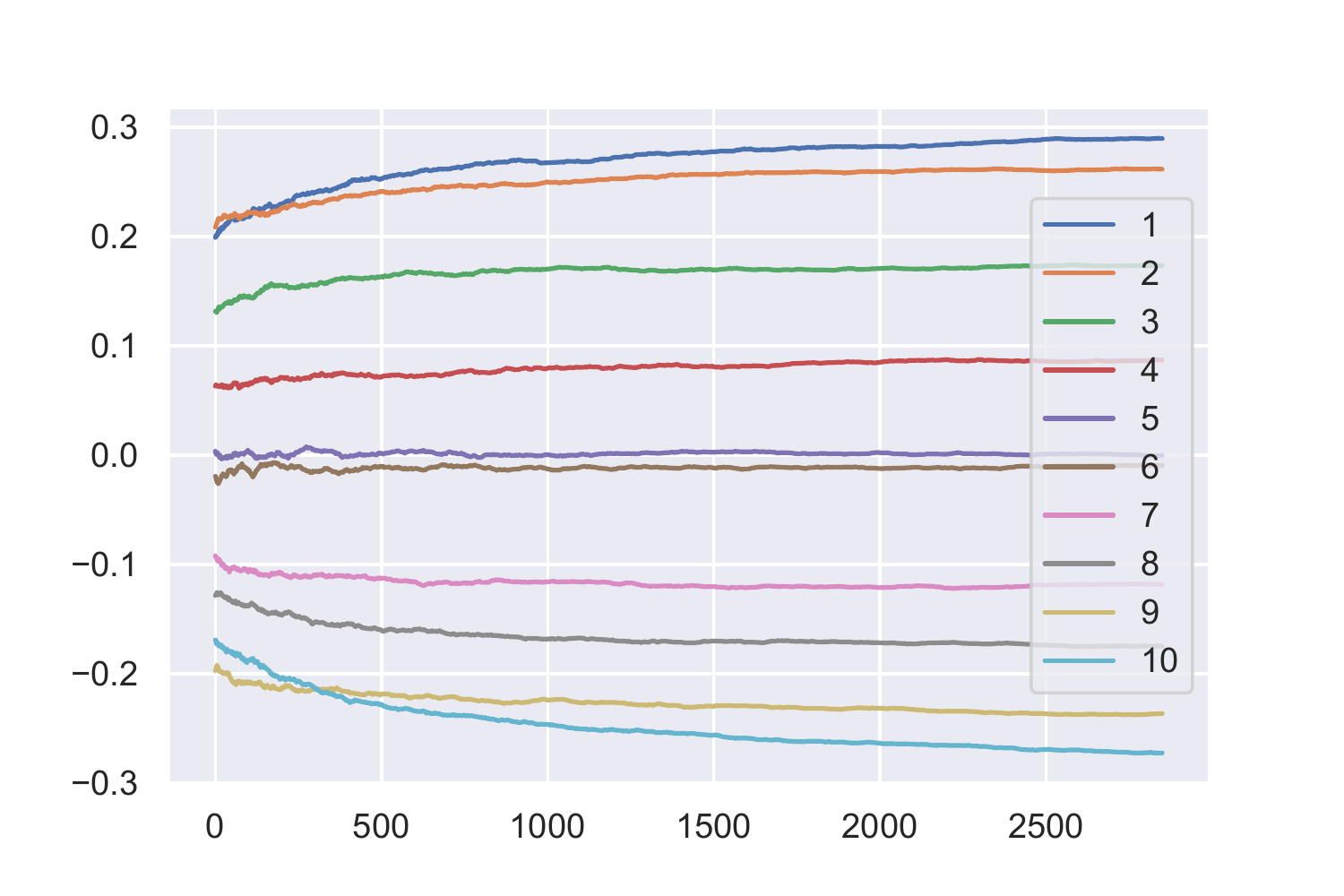}
  \caption{Online updates of player ratings $\epsilon_i^t$ over time in a transitive game with $n=10$ players sorted by decreasing strength, i.e. $P_{ij}>0$ when $i<j$. The ratings are averaged over 1000 simulations of 3000 timesteps each. For each timestep of each simulation, a pair of players is chosen at random to play each other, and the outcome of the game is sampled at random according to the probabilities $\widetilde{P}$. In the Hyperbolic Elo case, the empirical averages $\widetilde{P}^t_{ij}$ for a given simulation are computed using data from that simulation only. (Top) Elo update rule; (Middle) Hyperbolic Elo update rule, $g(x)=x$; (Bottom) Hyperbolic Elo update rule, $g(x)=\varphi_\beta(1)(x-\frac{1}{2})$. x-axis: time $t$. y-axis: player ratings. Note that contrary to the Elo case, Hyperbolic Elo ($\beta=5$) correctly ranks the players (players 1, 2, 9, 10 are incorrectly ranked in the Elo case). We check that the player ratings - in both Elo and Hyperbolic Elo cases - converge with good accuracy to the ratings found by minimizing the binary cross-entropy loss in Definition~\ref{elodef}. In accordance with the theory of Stochastic Approximation, we take in all cases a learning rate $\eta_t= \frac{32}{t^{0.8}}$, so that it satisfies the Robbins-Monro conditions. The first 150 timesteps out of 3000 have been removed to make the figure more readable.}
  \label{elo_online}
\end{figure}

\clearpage 

\begin{table}
  \caption{Average sign accuracy in $\%$ over 3 seeds and game sizes $n=50,75,100$. Results are presented as: "overall (train, test)".}
  \label{tab2}
  \centering
  \begin{tabular}{lllll}
    \toprule
     Game & Elo  & $m$-Elo  & NormalD & Ours \\
    \midrule
    connect four & 86 (87, 86) &   94 (94, 89)  &  94 (95, 89)    &    \textbf{97} (99, 85)\\
    5,3-Blotto  &71 (73, 54) &  \textbf{99} (100, 94)& 99 (100, 88) & \textbf{99} (100, 94) \\
    tic tac toe &93 (93, 92) &  96 (96, 91)  &  96 (97, 92) & \textbf{98} (100, 85) \\
    Kuhn-poker &81 (81, 80)  &  91 (91, 90) &  92 (92, 91) & \textbf{96} (98, 84) \\
    AlphaStar &86 (87, 85) &  92 (93, 87) & 92 (92, 88) & \textbf{95} (96, 85)\\
    quoridor(board size 4) & 87 (87, 83)& 92 (93, 84) & 93 (94, 86) &      \textbf{96} (98, 80)\\
    Blotto & 77 (77, 75)  & 94 (95, 91) & \textbf{95} (95, 91)  & \textbf{95} (97, 85) \\
    go(board size 4) & 84 (84, 80) &  93 (94, 85) & 93 (94, 86) &    \textbf{97} (99, 79) \\
    hex(board size 3) & 93 (93, 93)  &  96 (97, 89) &  97 (98, 91)   &   \textbf{98} (99, 85)\\
    \bottomrule
  \end{tabular}
\end{table}

\begin{table}
  \caption{StDev of sign accuracy in $\%$ over 3 seeds and game sizes $n=50, 75, 100$ related to Table~\ref{tab2}. Results are presented as: "overall (train, test)".}
  \label{tab3}
  \centering
  \begin{tabular}{lllll}
    \toprule
     Game & Elo  & $m$-Elo  & NormalD & Ours \\
    \midrule
    connect four & 1.4 (1.3, 4.3) & 0.9 (0.9, 3.4) &  0.8 (0.8, 5.2)  & 0.8 (1, 3.2) \\
    5,3-Blotto  & 1.1 (2.7, 11) &  0.8 (0, 8) & 0.4 (0, 3.1)& 1.2 (0, 8.8) \\
    tic tac toe & 0.9 (0.8, 2.7) &  0.6 (0.7, 2.2) &  0.7 (0.7, 1.4) & 0.5 (0.3, 3.7)\\
    Kuhn-poker & 0.1 (0.4, 3.2)  &  0.6 (0.8, 2.6) &  0.6 (0.7, 1.8) & 1.2 (1.1, 3.5) \\
    AlphaStar & 1.8 (1.7, 3.2) &  1.1 (1.2, 2.1) & 0.9 (0.9, 1.9) & 1.1 (1.3, 2.2)\\
    quoridor(board size 4) & 1.7 (1.7, 2) & 0.8 (1, 2.9) & 1 (1.2, 1.6) &      0.8 (1, 6.4)\\
    Blotto & 1.1 (1.2, 2.2) & 0.9 (1, 1.9) & 0.9 (0.9, 2.6) & 1.5 (1.6, 4.9)\\
    go(board size 4) & 2 (2.1, 1.5) & 1.7 (1.8, 3.2) & 2 (2.2, 3.4) & 0.8 (1, 4.3) \\
    hex(board size 3) & 1.1 (1.1, 1.7)  &  0.8 (0.9, 4.7) & 1 (0.9, 4.2) &   0.8 (0.8, 5)\\
    \bottomrule
  \end{tabular}
\end{table}

\begin{table}
  \caption{Average MAE ($\times 100$) over 3 seeds and game sizes $n=50, 75, 100$. Results are presented as: "overall (train, test)".}
  \label{tabmae}
  \centering
  \begin{tabular}{lllll}
    \toprule
     Game & Elo  & $m$-Elo  & NormalD & Ours \\
    \midrule
    connect four & 22 (22, 23) &  16 (15, 20) & 17 (17, 21) & \textbf{5} (2, 27)\\
    5,3-Blotto  & 24 (24, 31) & 7 (6, 20) & 9 (8, 22)   &   \textbf{2} (0, 19)\\
    tic tac toe & 18 (17, 19) & 14 (14, 19)  & 15 (14, 19)  &    \textbf{4} (1, 25) \\
    Kuhn-poker & 7 (7, 7) & \textbf{3} (3, 4) & \textbf{3} (3, 4) & \textbf{3} (2, 8) \\
    AlphaStar & 14 (14, 14) & 9 (9, 11) & 10 (10, 12) & \textbf{6} (4, 18)\\
    quoridor(board size 4) & 14 (14, 14) & 11 (11, 15) & 11 (11, 15)  &  \textbf{4} (2, 21)\\
    Blotto & 12 (12, 13) & 7 (7, 9)  & 7 (7, 9) & \textbf{5} (4, 14) \\
    go(board size 4) & 18 (18, 20) & 15 (14, 19) & 16 (15, 20)  &    \textbf{4} (2, 26)\\
    hex(board size 3) & 20 (20, 21) & 16 (16, 21) & 16 (16, 21)   &   \textbf{4} (1, 27)\\
    \bottomrule
  \end{tabular}
\end{table}

\begin{table}
  \caption{StDev of MAE ($\times 100$) over 3 seeds and game sizes $n=50,75,100$ associated to Table~\ref{tabmae}. Results are presented as: "overall (train, test)".}
  \label{tabmaes}
  \centering
  \begin{tabular}{lllll}
    \toprule
     Game & Elo  & $m$-Elo  & NormalD & Ours \\
    \midrule
    connect four & 0.6 (0.5, 2.2) &  1.7 (1.8, 2.3) & 1.2 (1.3, 2.8)& 1.2 (1.3 1.9)\\
    5,3-Blotto  & 0.1 (0.4, 4.3) & 0.8 (0.4,4.1) & 0.4 (0.3, 3)  & 0.5 (0.1, 5.2) \\
    tic tac toe & 0.7 (0.8, 1.1) & 1 (1.1, 1.1) & 0.8 (0.8,1.4)  & 0.4 (0.4, 1.7) \\
    Kuhn-poker & 0 (0, 0.3) & 0.1 (0.1, 0.2) & 0.1 (0.1, 0.2) & 0.5 (0.4, 1) \\
    AlphaStar & 1.3 (1.3, 1.1) & 0.9 (0.9, 0.6) & 0.7 (0.8, 0.7) & 0.5 (0.5, 1.1)\\
    quoridor(board size 4) & 0.5 (0.5, 0.7) & 0.8 (0.8, 1.3) & 0.6 (0.7, 1) & 0.5 (0.7, 2.2)\\
    Blotto & 0.2 (0.2, 0.3) & 0.4 (0.4, 0.3) & 0.4 (0.4, 0.3) & 0.8 (1, 2) \\
    go(board size 4) & 1 (1.1, 1) & 1.3 (1.4, 1.2) & 1 (1.1, 1.3) & 0.6 (0.7, 1.5) \\
    hex(board size 3) & 0.5 (0.4, 1.3) & 0.7 (0.9, 1.5) & 0.7 (0.7, 1.4) &   0.6 (0.7, 1.3)\\
    \bottomrule
  \end{tabular}
\end{table}

\end{document}